# Detection and Signal Processing for Near-Field Nanoscale Fourier Transform Infrared Spectroscopy

*Jonathan M. Larson [‡] (ORCID: 0000-0002-5389-0794), Hans A. Bechtel [⊥],[*] (ORCID: 0000-0002-7606-9333), and Robert Kostecki [‡],[*] (ORCID: 0000-0002-4014-8232)*

[‡] Energy Storage & Distributed Resources Division, Lawrence Berkeley National Laboratory, Berkeley, California, 94720, United States

[⊥] Advanced Light Source, Lawrence Berkeley National Laboratory, Berkeley, California, 94720, United States

[*]Corresponding Authors: Hans A. Bechtel (habechtel@lbl.gov) and Robert Kostecki (R_Kostecki@lbl.gov)

## ABSTRACT

Researchers from a broad spectrum of scientific and engineering disciplines are increasingly using near-field infrared spectroscopic techniques to characterize materials nondestructively and with nanoscale spatial resolution. However, sub-optimal understanding of a technique's implementation can complicate data interpretation and even act as a barrier to enter the field. Here we outline the key detection and processing steps involved in producing scattering-type near-field nanoscale Fourier transform infrared spectra (nano-FTIR). The largely self-contained work (i) explains how normalized complex-valued nano-FTIR spectra are generated, (ii) rationalizes how the real and imaginary components of spectra relate to dispersion and absorption respectively, (iii) derives a new and generally valid equation for spectra which can be used as a springboard for additional modeling of the scattering processes, and (iv) provides an algebraic expression that can be used to extract the sample's local extinction coefficient from nano-FTIR. The algebraic expression is validated with nano-FTIR and attenuated total reflectance Fourier transform infrared (ATR-FTIR) spectra on samples of polystyrene and Kapton.

## TEXT

### 1. Introduction

For at least half a century, Fourier transform infrared spectroscopy (FTIR) has been regarded as a gold standard for nondestructive chemical and structural fingerprinting.[1-3] This is due to the relatively low energy of infrared (IR) light, and the technique's sensitivity to changing electric dipole moments, such as those in molecular and crystal lattice vibrations. Moreover, the vast majority of materials are IR active and possess a unique IR spectrum signature, thus FTIR is ubiquitous in both academia and industry. Unfortunately, because of the relatively long wavelengths of infrared light and related diffraction limit, the spatial resolution for FTIR has been limited to *ca.* 1 - 1,000 $\mu m$.[4-7] Thus, FTIR's utilization during the so-called "nano-revolution"[8] during the last *ca.* 35 years has played an insignificant role in the characterization of nanostructures and associated nanoscale phenomena due to its inadequate spatial resolution. However, over the last decade, with the coalescence of scattering-type, scanning near-field optical microscopy (s-SNOM),[9] high power broadband IR sources, IR interferometry, and lock-in amplification techniques, scattering-type near-field nanoscale Fourier transform IR spectroscopy (nano-FTIR) has been realized,[10-17] commercialized, and its utilization has grown significantly in recent years.

The broad appeal and impact of nano-FTIR as a tool for sensing local IR-active physicochemical phenomena has been demonstrated by a large number of studies that span the gamut of science, technology, and engineering. A non-exhaustive subset of the many and diverse focus topics are nanoconfinement in metal-organic frameworks,[18] probing interphases between hard and soft condensed matter,[19] mapping catalytic reactions,[20] antenna assisted characterization of dielectric anisotropy,[21] phonon polaritons,[22,23] insulator-metal transitions,[24] mapping strain in 2D materials,[25] nanoscale dielectric properties,[26,27] organic matter contents of meteorites,[28] spectroscopic nanoimaging,[29] nonequilibrium physics,[30] battery electrode materials,[31] electrochemical interfaces and interphases,[32] and biological studies of proteins,[33,34] viruses,[35] linkers for biomolecule targeting,[36] living cells,[37] and spike proteins on COVID-19 viruses.[38]

Unfortunately, the detection scheme that underpins and enables the experimental realization of nano-FTIR is nontrivial. This methodological complexity, coupled with the diversity in practitioner-field expertise, creates an environment ripe for theoretical and conceptual barriers to robust understanding of (i) the detection and signal processing steps that generate and explain the complex-valued nano-FTIR spectrum, and (ii) the electrodynamic scattering process that takes place at the tip-sample region. There have been a number of publications addressing the later,[26,39-43] but a need persists for a rigorous and self-contained work that addresses the former – this is the aim of the present work. In order to avoid potential confusion between nano-FTIR and techniques that have similar purposes and abbreviations, we note that photothermal approaches for IR nano-spectroscopy have also been recently pioneered[44-47] (e.g. "nanoIR" or "AFM-IR"). These techniques are distinct in implementation from nano-FTIR, and are not the subject of this work.

## 2. Results and Discussion

*Experimental Setup and Data Acquisition Summary.* To provide a qualitative explanation for each of the essential steps involved with nano-FTIR acquisition, and identify key experimental components along the way; Fig. 1 serves as a critical visual aid and reference throughout the paper. First, a metal-coated atomic force microscope (AFM) probe oscillating at *ca.* $10^2$ kHz (period $T_c = 10 \, \mu s$ and tip-sample distance $d(t)$) is brought into tapping mode[48] engagement with a substrate of interest (Fig. 1) and positioned above a location of interest. Broadband IR light is introduced (solid red arrow on left-hand side of Fig. 1) into a portion of the experimental setup whose combined constituents function as an asymmetric Michelson interferometer[14,49] (AMI, items within gray box in Fig. 1). The asymmetric terminology speaks to the difference with respect to symmetric interferometers, which are typically used in conventional FTIR. Symmetric interferometers situate the sample of interest after the interferometer, while in asymmetric setups the sample is placed within, and is part of, the interferometer itself. As will be shown in detail later, a key advantage to the asymmetric set up is that mathematical (Fourier) transformation of data collected yields both optical amplitude and phase (or real and imaginary) information that is used to quantify sample properties.

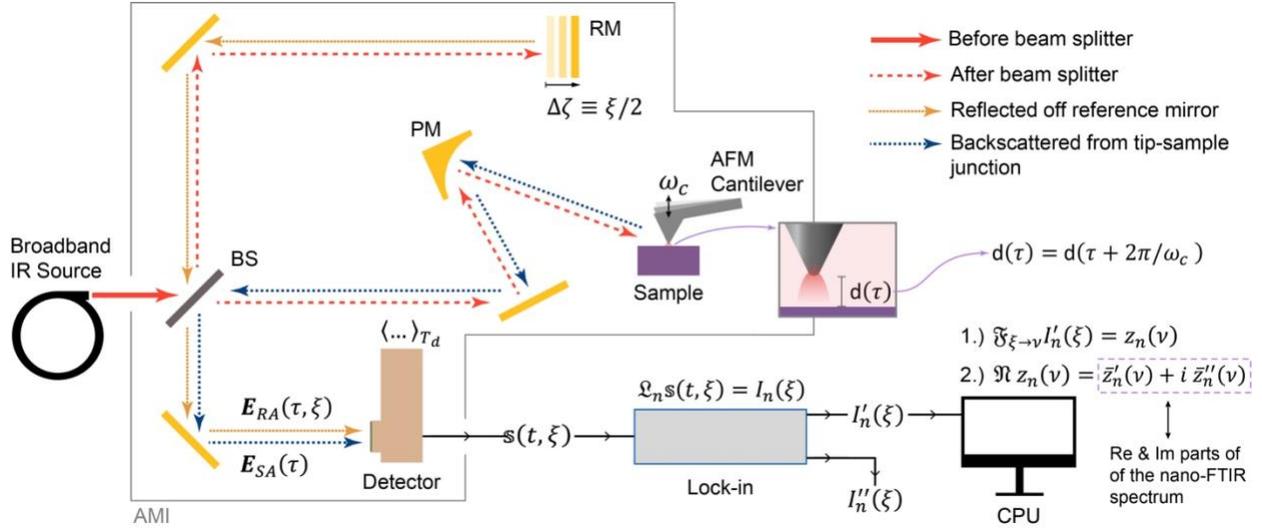

**Fig. 1.** Schematic of the experimental setup and processing steps for scattering scanning near-field nanoscale Fourier transform infrared spectroscopy. Operator notation that is used in the rest of the text matches the diagram: detector time averaging ($\langle...\rangle_{T_d}$), lock-in amplification ($\mathfrak{L}$), Fourier transformation ($\mathfrak{F}$), and normalization ($\mathfrak{N}$).

The broadband IR light source is typically produced via difference frequency generation (DFG) with a fiber-laser[11] (broadband) or a synchrotron[12,16,17] (ultrabroadband), although other sources such as high temperature plasmas are also available.[50] The typical detectable bandwidths of these two sources are shown in Fig. 2. As a quick aside, we note that when the light source is a broadband IR laser, nano-FTIR is the commonly used acronym. However, when the light source is an ultrabroadband synchrotron, SINS (synchrotron infrared nanospectroscopy) is the commonly used acronym. In both cases however, the signal processing and detection scheme is the same. Herein, we use the nano-FTIR acronym as encompassing both these approaches.

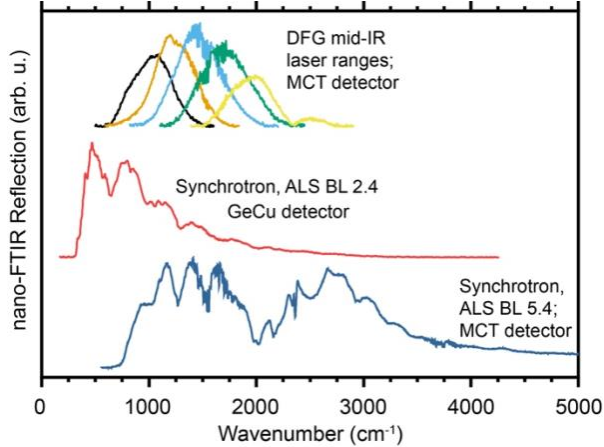

**Fig. 2.** Raw, unnormalized, nano-FTIR reflection spectra of Au or Si for selected broadband infrared light sources. The MCT detector data sets with a DFG mid-IR laser source (top-most spectra, multiple colors) utilized a ZnSe beamsplitter. The GeCu detector data from ALS BL 2.4 (red trace) was collected with a KRS-5 beamsplitter and the MCT detector data from ALS BL 5.4 (dark blue trace) was collected with a KBr beamsplitter.

As the broadband light enters the AMI (solid red arrow on left-hand side of Fig. 1), it first passes through a beam splitter (BS in Fig. 1). As the name implies, the BS splits the beam in two, sending each half in a different direction (see red dashed arrows in Fig. 1). Half the power is reflected, and travels toward a reference mirror (RM, top middle Fig. 1), while the other half is transmitted and travels toward the AFM tip-sample region. These unique beam paths are referred to as the "reference arm" and "sample arm" of the AMI, respectively. Both arms return light (dotted yellow and blue arrows in Fig. 1), which is recombined and measured at the detector (bottom right of AMI in Fig. 1).

Within the reference arm, as illustrated in Fig. 1, the light travels to (red dashed arrows) and reflects off (yellow dotted arrows) the RM, then transmits through the BS, before combining with light from the sample arm and being detected. It is important to make clear that the RM is movable. In particular, during nano-FTIR data acquisition, the RM will linearly translate, changing the path length difference between the sample and reference arms. During this translation, the sample and reference arm path lengths should be approximately similar, and exactly similar for a brief part of the scan; a position known as the zero path difference, or ZPD. Herein, we assign a generalized coordinate, $\zeta$, to identify the spatial position of the RM. Any change in the RM position by some $\Delta\zeta$ (as shown in Fig. 1) will change the distance the reference light will need to travel in the reference arm before detection. Specifically, the change in reference beam travel distance before detection will be $2\Delta\zeta$, where the factor of two arises from light traveling both to, and from, the RM. Thus, translating the RM amounts to shifting the reference beam in space by an amount $2\Delta\zeta$, say $\xi$, and therefore the total signal detected will be a function of $\xi$.

The light entering the sample arm of the AMI is focused with a parabolic mirror (PM in Fig. 1) toward the tip-sample region. A fraction of this incident light is backscattered along the same beam path from where it came (see blue dotted arrows in Fig. 1). Eventually, it recombines with the reference light, and is subsequently detected. The light backscattered from the region around the sample and vertically oscillating (Fig 3(a)) metallic AFM probe can be considered a linear combination of two distinct fields, classified by the "kind" of light that drives the scattering. The first kind, and most intuitive, is the conventional diffraction-limited incident IR that bathes the entire tip-sample region in "far-field" light (red arrows and shading in Fig. 3(d) and (e)). The second kind, is the diffraction-limit-breaking "near-field" – induced by the incident IR in combination with probe conductivity and geometry – which is both enhanced and spatially confined in comparison to the incident IR[26,39,41-43] (darker red shading in Fig. 3(f) and (g)). The former drives "far-field scattering" from the sample and AFM probe shank (Fig. 3(d) and (e)), while the later drives "near-field scattering" from nanoscopic volume around the probe tip end (Fig. 3(f) and (g)). Because of this, as will be discussed in more detail later, the relevant electric field component of light backscattered from the tip-sample region will be described as a linear combination of these two contributors: $E_{SA} = E^{FF} + E^{NF}$, where $SA$, $FF$, and $NF$ refer to sample arm, far-field, and near-field, respectively.

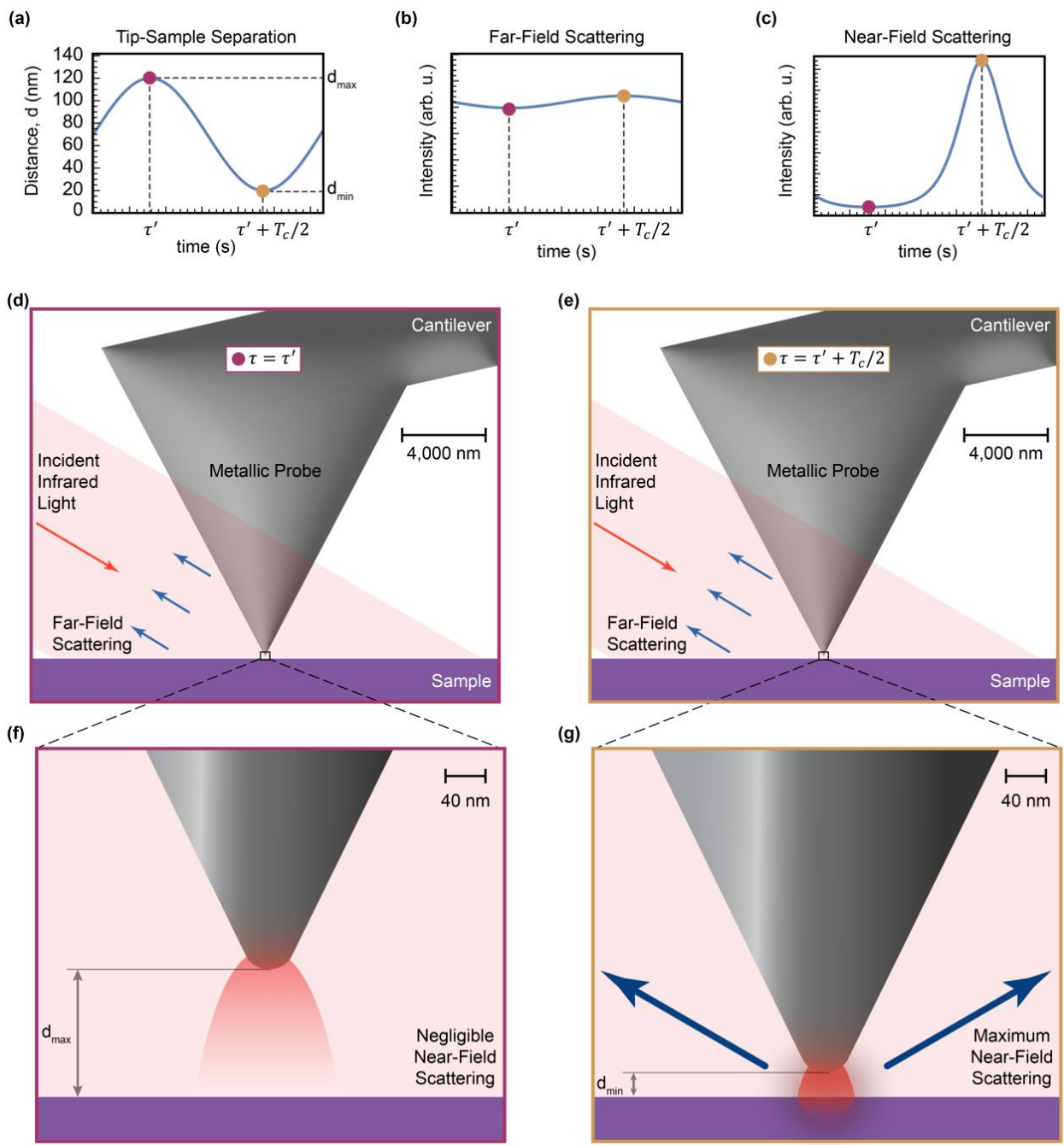

**Fig. 3.** Representative semi-quantitative plots of the typical (a) tip-sample separation distance[43], (b) far-field scattering intensity, and (c) near-field scattering intensity as a function of time, over one cantilever period, $T_c$. Key points in time associated with the greatest ($\tau = \tau'$) and least ($\tau = \tau' + T_c/2$) tip-sample separation are highlighted with purple and gold circles. (d) Microscale illustration of the tip-sample system, bathed in incident IR light, at the point in time of greatest, $\tau = \tau'$, ((e) least, $\tau = \tau' + T_c/2$) tip-sample separation. Note that the far-field scattering is only slightly modified by the nanoscale geometrical perturbations associated with cantilever oscillation. Panels (f) and (g) provide nanoscale Zoom-in illustrations of the previous two panels and emphasize how near-field scattering events from nanoscopic volumes below the probe tip end are highly dependent on the nanoscale geometrical perturbations associated with cantilever oscillation.

As the AFM cantilever oscillates, far-field scattering from the sample and probe shank, $E_{FF}$, is only modestly altered (Fig. 3(b), (d), and (e)), while near-field scattering, $E_{NF}$, dramatically changes with the tip-sample separation distance (Fig. 3(c), (f), and (g)). In particular, the enhanced near-field strongly interacts with the sample at the points in time of least tip-sample distance (Fig. 3(a), (c), (f), and (g)), causing photons to be periodically scattered from the narrow region of the sample located under the tip's end. This small portion of the total backscattered light is attributable to scattering processes occurring within nanoscopic volumes around the probe tip end and carries local materials properties encoded within it. Therefore, the entire aim/goal of the detection and signal processing scheme is to extract this localized information.

In the following section, we present physics-based continuum mathematical models for the detection and signal processing steps, which ultimately extract meaningful information from the signal sensed by the detector, as a function of time and reference mirror position. We first provide the order-of-magnitude timescales at play (the periods) in the experimental system: IR field oscillations ($T_{IR} = 10^{-14}s$), detector time resolution ($T_d = 10^{-6}s$), cantilever oscillation ($T_c = 10^{-5}s$), and the lock-in time averaging period ($T_L = 10^{-4}s$). Secondly, we mention how variable independence between $t$ and $\xi$ is practically attained. There are two ways that ensure such independence: collecting time-dependent data at a fixed mirror position, or by moving the mirror significantly slower than any other relevant timescale. Regardless of how this is technically implemented, $t$ and $\xi$ are treated as independent variables herein.

***Detector Time-Averaging $\langle ... \rangle_{T_d}$ and Mathematical Framework.*** The infrared detector, usually a liquid nitrogen cooled mercury cadmium telluride (MCT) (or other, e.g., a liquid helium cooled germanium copper (GeCu) detector), measures time-averaged electromagnetic energy transferred from the Poynting vector to the photodetector (Fig. 4). The time-averaging period is dictated by hardware, and as mentioned above, is typically $T_d = 10^{-6}s = 1\ \mu s$. Thus, the magnitude of the detector signal depends on a time average, over $T_d$, of the Poynting vector at the detector surface: $\mathbb{s}(t,\xi) = \langle \eta \int \boldsymbol{S}(\boldsymbol{x}_d, \tau, \xi) \cdot \hat{\boldsymbol{x}}\, da \rangle_{T_d}$. Here, $\mathbb{s}(t,\xi)$ is the real-valued signal output from the detector, $\eta$ is some constant for energy conversion between the field (just outside the detector surface) and the photodetector (at $x = 0$), $\boldsymbol{S}(\boldsymbol{x}_d, \tau, \xi)$ is the Poynting vector at the detector surface, $\hat{\boldsymbol{x}}$ is the unit vector inward and normal to the detector surface, the brackets $\langle ... \rangle_{T_d}$ indicate a time-average over $T_d$, the $d$ subscript on $\boldsymbol{x}_d$ is used to emphasize that the integral is over the surface area of the detector: $\boldsymbol{x}_d = (0, y_d, z_d)$, and the time variables are related by $t = \tau - T_d/2$. Figure 4 schematically aids in visualizing the coordinate system and described processes.

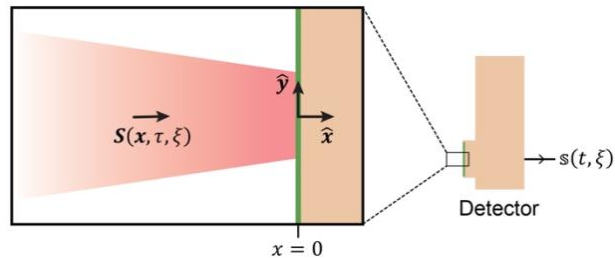

**Fig. 4.** Schematic zoom-in of the detector surface with incident IR light.

While not pictured in Fig. 1, light from the reference and sample arms of the AMI (dotted yellow and blue arrows in Fig. 1) is focused onto the detector with optics including another parabolic mirror. This results in the light beam incident on the detector surface having a cone-like geometry, as illustrated in Fig. 4. In this work, we approximate the Poynting flux directionality at the detector surface as being solely normally incident ($\hat{x}$ direction) to the detector surface – akin to considering the spatial average of the Poynting flux about its azimuthally symmetric beam path. Thus, we rewrite $S(x_d, \tau, \xi)$ as $S(\tau, \xi)\hat{x}$, and the surface integral simplifies to the product of the detector area, $a$, and the Poynting flux evaluated at any point on the surface of the detector: $\mathbb{s}(t, \xi) = \langle \eta a S(\tau, \xi) \rangle_{T_d}$. Furthermore, because $S = |\mathbf{S}| = (1/\mu_0)|\mathbf{E} \times \mathbf{B}| = (1/\mu_0)|\mathbf{E}||\mathbf{B}|$, and $|\mathbf{E}| = c|\mathbf{B}|$ in source free regions, like the region directly adjacent to the detector, the expression for the detector signal becomes

$$\mathbb{s}(t, \xi) = \frac{\eta a}{c \mu_0 T_d} \int_{t - \frac{T_d}{2}}^{t + \frac{T_d}{2}} |\mathbf{E}(\tau, \xi)|^2 \, d\tau + n(t). \qquad (1)$$

In Equation (1) $c$ and $\mu_0$ are the speed of light and permeability of free space, respectively. Furthermore, the electric field, $\mathbf{E}$, is real-valued, and we have additionally introduced a term, $n(t)$, which serves to represent randomized noise implicit to detection or from ambient light. As described above, the electric field, $\mathbf{E}$, is comprised of a sum of non-trivial broadband fields, one from the sample arm of the AMI, $\mathbf{E}_{SA}$, and one from the reference arm of the AMI, $\mathbf{E}_{RA}$. The square of the field is $|\mathbf{E}|^2 = \mathbf{E} \cdot \mathbf{E} = (\mathbf{E}_{SA} + \mathbf{E}_{RA}) \cdot (\mathbf{E}_{SA} + \mathbf{E}_{RA}) = \mathbf{E}_{SA} \cdot \mathbf{E}_{SA} + 2\mathbf{E}_{SA} \cdot \mathbf{E}_{RA} + \mathbf{E}_{RA} \cdot \mathbf{E}_{RA}$. Moreover, because we are detecting light far from the sources of radiation, $\mathbf{E}_{SA}$ and $\mathbf{E}_{RA}$ are considered plane waves with polarization orthogonal to $\mathbf{S}$.[51] In addition, for modeling simplicity, we treat all fields incident on the detector as being similarly polarized, say in the $\hat{y}$ direction. Thus, at the detector surface we can write, $\mathbf{E}_{SA} = E_{SA,y}\hat{y}$ and $\mathbf{E}_{RA} = E_{RA,y}\hat{y}$. Utilizing all the above results in the expression for the detector signal yields

$$\mathbb{s}(t, \xi) = \frac{\eta a}{c \mu_0 T_d} \int_{t - \frac{T_d}{2}}^{t + \frac{T_d}{2}} \left[ E_{SA,y}^2(\tau) + 2 E_{SA,y}(\tau) E_{RA,y}(\tau, \xi) + E_{RA,y}^2(\tau, \xi) \right] d\tau + n(t). \qquad (2)$$

We point out that time dependence of the fields enters in two ways: field oscillations and periodic changes in the backscattered light from the sample arm (due to geometric changes at the tip-sample region from cantilever oscillations). Thus, the later time dependence arises only in $E_{SA,y}$, while the time dependence in $E_{RA,y}$ terms come solely from field oscillations. This difference is relevant, and will be exploited later.

Moving forward with the calculation, by utilizing an inverse Fourier transform in time, we will re-write the occurrences of real-valued field components $E_{SA,y}$ and $E_{RA,y}$ in Equation (2), as generalized multichromatic fields which satisfy the Maxwell and wave equations in free space. Such an expression, as described elsewhere,[52] takes the form

$$f(t,x) = \frac{1}{\sqrt{2\pi}} \int_{-\infty}^{\infty} \tilde{f}(\omega) e^{i\omega\left(\frac{x}{c}-t\right)} d\omega, \tag{3}$$

where $\tilde{f}(\omega)e^{i\omega x/c} = \tilde{f}(\omega,x) = \mathfrak{F}_{t \to \omega} f(t,x)$ is the Fourier transform in time of $f(t,x)$, and $\tilde{f}(\omega) = \tilde{f}(\omega,0) = \mathfrak{F}_{t \to \omega} f(t,0)$ is possibly complex. Using this functional form and the fact that the detector is located at $x = 0$, without loss of generality, $E_{SA,y}$ becomes

$$E_{SA,y}(\tau) = \frac{1}{\sqrt{2\pi}} \int_{-\infty}^{\infty} \tilde{E}_{SA,y}(\omega) e^{-i\omega\tau} d\omega. \tag{4}$$

As for the field from the reference arm, $E_{RA,y}$, we account for spatial shifts arising from various RM positions, with the following replacement in Equation (3), $x \to x + \xi = 0 + \xi$, giving

$$E_{RA,y}(\tau,\xi) = \frac{1}{\sqrt{2\pi}} \int_{-\infty}^{\infty} \tilde{E}_{RA,y}(\omega) e^{i\omega\left(\frac{\xi}{c}-\tau\right)} d\omega. \tag{5}$$

Note that both $\tilde{E}_{SA,y}$ and $\tilde{E}_{RA,y}$ in equations (4) and (5) are generally complex.

Inserting the above relations for $E_{SA,y}(\tau)$ and $E_{RA,y}(\tau,\xi)$ into Equation (2), and switching order of integration gives

$$\mathbb{s}(t,\xi) = \frac{\eta a}{2\pi c\mu_0}\Bigg[\int_{-\infty}^{\infty}\int_{-\infty}^{\infty} \tilde{E}_{SA,y}(\omega_1)\tilde{E}_{SA,y}(\omega_2) \left\{\frac{1}{T_d}\int_{t-\frac{T_d}{2}}^{t+\frac{T_d}{2}} e^{-i\tau(\omega_1+\omega_2)}d\tau\right\} d\omega_1\, d\omega_2 +$$

$$2\int_{-\infty}^{\infty}\int_{-\infty}^{\infty} \tilde{E}_{SA,y}(\omega_3)\tilde{E}_{RA,y}(\omega_4) e^{i\xi\frac{\omega_4}{c}} \left\{\frac{1}{T_d}\int_{t-\frac{T_d}{2}}^{t+\frac{T_d}{2}} e^{-i\tau(\omega_3+\omega_4)}d\tau\right\} d\omega_3 d\omega_4 +$$

$$\int_{-\infty}^{\infty}\int_{-\infty}^{\infty} \tilde{E}_{RA,y}(\omega_5)\tilde{E}_{RA,y}(\omega_6) e^{i\xi\left(\frac{\omega_5}{c}+\frac{\omega_6}{c}\right)} \left\{\frac{1}{T_d}\int_{t-\frac{T_d}{2}}^{t+\frac{T_d}{2}} e^{-i\tau(\omega_5+\omega_6)}d\tau\right\} d\omega_5 d\omega_6\Bigg] + n(t), \tag{6}$$

where the numeric subscripts on the $\omega$'s serve to indicate they are simply integration variables. The time average integrals enclosed within curly brackets in Equation (6) have an analytic solution of a product of exponential and sinc functions: $\exp[-it\Omega_{lm}]\text{sinc}[T_d\Omega_{lm}/2]$, where $\Omega_{lm} = \omega_l + \omega_m$. In the case of the time average integral in the third term, it further reduces to an effective

Dirac delta distribution, as described in the Supporting Information. Incorporating these results into the above yields,

$$s(t,\xi) = \mathcal{E}_{SS}(t) + \mathcal{E}_{SR}(t,\xi) + \mathcal{E}_{RR} + n(t) \tag{7}$$

where,

$$\mathcal{E}_{SS}(t) = \frac{\eta a}{2\pi c \mu_0} \int_{-\infty}^{\infty} \int_{-\infty}^{\infty} \tilde{E}_{SA,y}(\omega_1) \tilde{E}_{SA,y}(\omega_2) \, \text{sinc}\left(\frac{T_d \Omega_{12}}{2}\right) e^{-i\Omega_{12} t} d\omega_1 \, d\omega_2,$$

$$\mathcal{E}_{SR}(t,\xi) = \frac{\eta a}{\pi c \mu_0} \int_{-\infty}^{\infty} \int_{-\infty}^{\infty} \tilde{E}_{SA,y}(\omega_3) \tilde{E}_{RA,y}(\omega_4) e^{i\xi \frac{\omega_4}{c}} \, \text{sinc}\left(\frac{T_d \Omega_{34}}{2}\right) e^{-i\Omega_{34} t} d\omega_3 \, d\omega_4,$$

and

$$\mathcal{E}_{RR} = \frac{\eta a}{2\pi c \mu_0} \int_{-\infty}^{\infty} |\tilde{E}_{RA,y}(\omega_6)|^2 d\omega_6,$$

and $\mathcal{E}_{SS}$, $\mathcal{E}_{SR}$, and $\mathcal{E}_{RR}$, are terms for the time-averaged electromagnetic energy detected and associated with various combinations of $E_{SA,y}$ and $E_{RA,y}$, as denoted by the two subscripts. The so-called "self-homodyne" signal associated with coupling/interference between far-field ($E^{FF}$) and near-field ($E^{NF}$) backscattered light is contained within $\mathcal{E}_{SS}(t)$ (because $E_{SA} = E^{FF} + E^{NF}$). The so-called "heterodyne" signal associated with coupling/inference between near-field backscattered light and light from the reference arm of the AMI is contained in $\mathcal{E}_{SR}(t,\xi)$.

***Detector Signal Expressed as Fourier Series in Time.*** With benefit of knowing that a lock-in operation is forthcoming, and that lock-in operations efficiently isolate Fourier expansion coefficients (see Supporting Information), it is advantageous to cast the detector signal as Fourier series. So, we proceed with this intention. The third term in Equation (7), $\mathcal{E}_{RR}$, is time independent as a result of detector time-averaging. However, the first and second terms, $\mathcal{E}_{SS}(t)$ and $\mathcal{E}_{SR}(t,\xi)$, maintain time dependence because geometrical changes associated with the oscillating AFM probe occur at order of magnitude longer time scales then the detector time averaging ($T_d \sim 1 \, \mu s$ whilst $T_c \sim 10 \, \mu s$). Under the reasonable assumption that the sample's local materials properties are constant – because the AFM is not in contact mode, and the IR photon energy is low – the geometrical changes are the only cause for detectible temporal changes in the electric field back-scattered from the IR-illuminated cantilever-sample system. Because the AFM cantilever is driven to oscillate with a known period, $T_c$, the geometrical configuration of the tip-sample system possesses the following temporal periodicity: $\boldsymbol{G}(t) = \boldsymbol{G}(t + T_c)$, where $\boldsymbol{G}(t)$ is the set of parameters that define a unique geometrical configuration. It follows that all remaining time-dependent signals will possess the same temporal periodicity: $\mathcal{E}_{SS}(t) = \mathcal{E}_{SS}(t + T_c)$ and $\mathcal{E}_{SR}(t,\xi) = \mathcal{E}_{SR}(t + T_c, \xi)$. Because $\mathcal{E}_{SS}(t)$ and $\mathcal{E}_{SR}(t,\xi)$ are real-valued and periodic in time, they can be represented as a series of orthonormal basis functions. Here we choose the cosine function with a phase, generally giving

$$\mathcal{E}_{SS}(t) = \sum_{\beta \geq 0}^{\infty} V_\beta \cos(\beta \omega_c t + \varphi_\beta) \tag{8}$$

and

$$\mathcal{E}_{SR}(t, \xi) = \sum_{n \geq 0}^{\infty} U_n(\xi) \cos(n\omega_c t + \theta_n(\xi)). \tag{9}$$

The motivation for casting $\mathcal{E}_{SS}(t)$ and $\mathcal{E}_{SR}(t, \xi)$ in such a formulation is due to the forthcoming lock-in operation, which will efficiently isolate and output expressions which are combinations of the expansion coefficients ($V_\beta, U_n(\xi)$) and/or phases ($\varphi_\beta, \theta_n(\xi)$) in Equations (8) and (9). For our modeling purposes then, we must derive expressions for the expansion coefficients in Equations (8) and (9) in terms of the fields embedded in Equation (7). However, it turns out, as will be made clear later, that for this work we only need to solve for $U_n(\xi)$. This is done by utilizing cosine orthogonality relations, and limiting cases of the sinc function, and is explicitly derived in the Supporting Information. The result is

$$U_n(\xi) = \frac{\eta a \, \text{sinc}\left(\frac{T_d n \omega_c}{2}\right)}{\pi c \mu_0 \cos(\theta_n(\xi))} \int_{-\infty}^{\infty} \left[\tilde{E}_{SA,y}^*(\omega - n\omega_c) + \tilde{E}_{SA,y}^*(\omega + n\omega_c)\right] \tilde{E}_{RA,y}(\omega) e^{i\xi \frac{\omega}{c}} d\omega. \tag{10}$$

Now, it is helpful to recall that the light backscattered from the tip-sample region (and measured at the IR detector) can be described as a sum of fields: $E_{SA,y}(t) = E_y^{FF}(t) + E_y^{NF}(t)$. That is, a sum of backscattered light due to far-field driven excitations ($E_y^{FF}$) and backscattered light due to near-field driven excitations ($E_y^{NF}$). Thus, occurrences of $\tilde{E}_{SA,y}(\omega)$ can be replaced with $\tilde{E}_y^{FF}(\omega) + \tilde{E}_y^{NF}(\omega)$. Doing so in Equation (10) reveals that every expansion coefficient, $U_n(\xi)$, is actually a sum of two terms, one related to far-field scattering, $u_n^{FF}(\xi)$, and one related to near-field scattering, $u_n^{NF}(\xi)$:

$$U_n(\xi) = u_n^{FF}(\xi) + u_n^{NF}(\xi), \tag{11}$$

where,

$$u_n^{FF}(\xi) = \frac{\eta a \, \text{sinc}\left(\frac{T_d n \omega_c}{2}\right)}{\pi c \mu_0 \cos(\theta_n(\xi))} \int_{-\infty}^{\infty} \left[\tilde{E}_y^{FF*}(\omega - n\omega_c) + \tilde{E}_y^{FF*}(\omega + n\omega_c)\right] \tilde{E}_{RA,y}(\omega) e^{i\xi \frac{\omega}{c}} d\omega,$$

and,

$$u_n^{NF}(\xi) = \frac{\eta a \, \text{sinc}\left(\frac{T_d n \omega_c}{2}\right)}{\pi c \mu_0 \cos(\theta_n(\xi))} \int_{-\infty}^{\infty} \left[\tilde{E}_y^{NF*}(\omega - n\omega_c) + \tilde{E}_y^{NF*}(\omega + n\omega_c)\right] \tilde{E}_{RA,y}(\omega) e^{i\xi\frac{\omega}{c}} \, d\omega.$$

Thus, $\mathcal{E}_{SR}(t,\xi)$ can either be described as one infinite sum with Fourier coefficients $U_n(\xi)$ (Equation (9)), or by using Equation (11) in (9), as two infinite sums,

$$\mathcal{E}_{SR}(t,\xi) = \sum_{n \geq 0} u_n^{FF}(\xi) \cos(n\omega_c t + \theta_n(\xi)) + \sum_{n \geq 0} u_n^{NF}(\xi) \cos(n\omega_c t + \theta_n(\xi)), \quad (12)$$

where the first (second) sum describes the detector time-averaged product of the reference field with far-field (near-field) backscattering. Therefore, via (11) and (12), it becomes clear that the $u_n^{NF}(\xi)$ coefficients comprise the heterodyne signal that is embedded in $\mathcal{E}_{SR}(t,\xi)$.

Furthermore, expansion coefficients $u_n^{FF}$, which define the time-dependence character of the far-field scattering, possess key attributes. As already described above, the dominant contributor to the total backscattered light is far-field scattering from microscale regions around the tip-sample region (Fig. 3(b), (d) and (e)). However, far-field scattering only modestly changes in time because the nanoscopic changes in probe position (from cantilever oscillation) are only weakly perturbative to the microscale geometry from which such scattering is occurring (Fig. 3(d) and (e)). Therefore, the zeroth and first harmonic Fourier coefficients, $u_0^{FF}$ and $u_1^{FF}$, which characterize a constant background and simple periodic change, should be necessary and (mostly) sufficient to describe the detector time-averaged product of the reference and far-field scattering, and so,

$$u_{n>1}^{FF} \cong 0. \quad (13)$$

***Lock-in Amplification $\mathfrak{L}$: Removal of Constant Terms and Time Dependence, Suppression of Noise, Isolation of Fourier Expansion Coefficients, and Formation of the Interferogram.*** Lock-in amplification is a process of performing two real-valued transformations on a polluted signal to extract information of interest. The polluted signal is typically a sum of a perfectly periodic function of interest, unwanted information, and noise. By time-averaging the product of the polluted signal and a cosine or sine function (with integer multiple of the frequency of interest), Fourier expansion coefficients that describe the embedded and obscured periodic function of interest can be isolated. This process typically also removes time dependence. It is common to think of the two real-valued transformations as one complex transformation that maps the single real signal to a complex signal who's real (imaginary) part is defined as the time average of the signal multiplied by cosine (sine). Because of this, it is common for lock-in amplifiers to output this defined complex signal. Following this convention, we herein denote the linear operator of the single complex lock-in operation as $\mathfrak{L}_m$, where $m$ is a positive integer that defines the "harmonic" being considered. A detailed accounting of all the essential mathematical steps that the lock-in effectively performs is provided in the Supporting Information.

The next step is to calculate $\mathfrak{L}_m \mathfrak{s}(t, \xi)$, where $\mathfrak{s}(t, \xi)$ is defined in Equations (7), (8), and (9). As the reference mirror translates, data is output from the lock-in and recorded as a function of $\xi$. The $\xi$-dependent real and imaginary outputs of the lock-in constitute the complex valued interferogram: $\mathfrak{L}_m \mathfrak{s}(t, \xi) \equiv I_m(\xi) = I'_m(\xi) + i\, I''_m(\xi)$. In passing, we caution the reader not to confuse the real and imaginary (or amplitude and phase) lock-in outputs, which define the complex valued interferogram, with the real and imaginary components (or the amplitude and phase) of the complex valued normalized nano-FTIR spectrum, which is still yet to be constructed.

As described in detail in the Supporting Information, the lock-in operation will cause two kinds of terms to vanish: those which are constant-in-time, and those which are randomized noise and non-monotonically increasing or decreasing. Therefore, as $\mathfrak{L}_m$ is applied to Equation (7), $\mathfrak{L}_m \mathcal{E}_{RR} = \mathfrak{L}_m n(t) = 0$, leaving $\mathfrak{L}_m \mathfrak{s}(t, \xi) = \mathfrak{L}_m \mathcal{E}_{SS}(t) + \mathfrak{L}_m \mathcal{E}_{SR}(t, \xi)$. Because $\mathcal{E}_{SS}(t)$ and $\mathcal{E}_{SR}(t, \xi)$ can be expressed in terms of an infinite sum of cosines (Equations (8) and (9)), and for details outlined in the Supporting Information, $\mathfrak{L}_m \mathfrak{s}(t, \xi)$ reduces to the following (where we have allowed the general index $m \to n$):

$$I_n(\xi) = \frac{V_n \cos(\varphi_n)}{2} + \frac{U_n(\xi) \cos(\theta_n(\xi))}{2} + i\left\{\frac{V_n \sin(\varphi_n)}{2} + \frac{U_n(\xi) \sin(\theta_n(\xi))}{2}\right\}. \quad (14)$$

Here, $I_n(\xi)$ is the time-independent complex valued interferogram of the $n^{th}$ harmonic. The real part of this quantity, $I'_n(\xi)$, is coined the interferogram of the $n^{th}$ harmonic, and is the critical $\xi$-dependent signal that is passed onto a computer for further processing (Fig. 1). As an example, interferograms of the 2nd harmonic for Si and polystyrene (PS) are provided in Figures 5a and 5b respectively. Before moving further along, we note in passing that the time-averaging period for the lock-in (see Supporting Information), $T_L = 10^{-4} s = 100 \mu s$, is an order of magnitude larger than the cantilever period; and so each lock-in output is a time average over roughly ten cantilever oscillations.

At this point, it is reasonable to pause and address a likely question: what is the difference between the complex interferograms of various harmonics (different values for $n$)? For reasons recently described above Equation (13), for the zeroth and fundamental harmonics, $U_{n \leq 1}(\xi)$ will be dominated by contributions from far-field scattering, $u_n^{FF}(\xi)$. Thus, $I_{n \leq 1}(\xi)$ is a complex interferogram with predominantly far-field scattering information encoded in it. On the other hand, when $n > 1$, far-field contributions become quite small (Equation (13)), $U_n(\xi) \to u_n^{NF}(\xi)$, and the complex interferogram, $I_{n>1}(\xi)$ has predominantly near-field scattering information encoded in it:

$$I_{n>1}(\xi) = \frac{V_n \cos(\varphi_n)}{2} + \frac{u_n^{NF}(\xi) \cos(\theta_n(\xi))}{2} + i\left\{\frac{V_n \sin(\varphi_n)}{2} + \frac{u_n^{NF}(\xi) \sin(\theta_n(\xi))}{2}\right\}. \quad (15)$$

For these reasons, nano-FTIR experiments focus on lock-in harmonics two and higher, so as to isolate near-field contributions. In the following expressions, we will do the same.

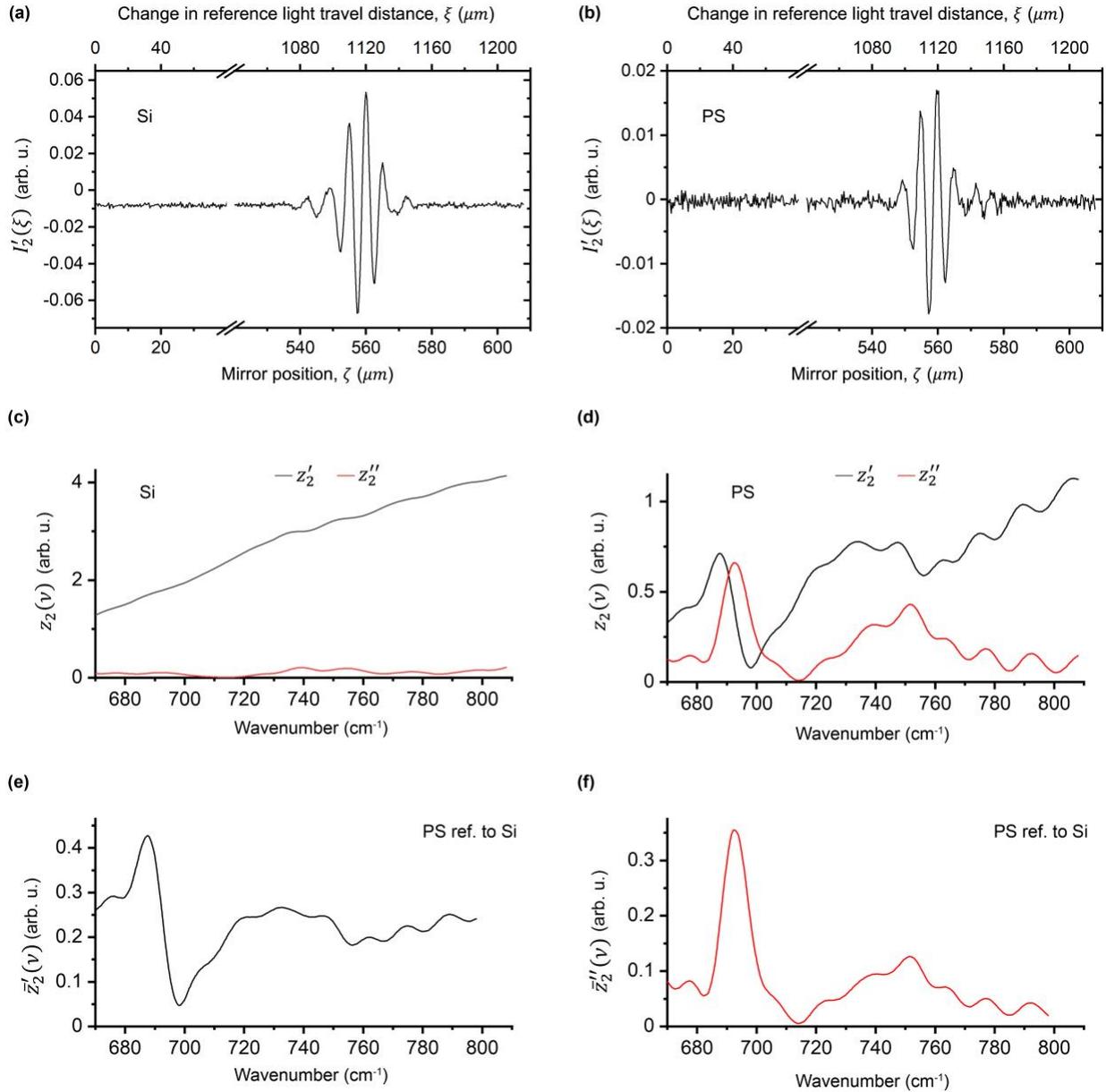

**Fig. 5.** Representative key data sets in the generation of a normalized complex valued nano-FTIR spectrum. Panels (a) and (b) are second harmonic nano-FTIR interferograms for Si and PS respectively, collected with a commercial Neaspec system utilizing a broadband IR laser and tapping amplitude of 85nm. Panels (c) and (d) are plots of the real ($z'_2$) and imaginary ($z''_2$) parts of the Fourier transforms of the interferograms ($z_2 = \mathfrak{F}_{\xi \to \nu} I'_2$) for Si and PS respectively. Panels (e) and (d) are, respectively, the real and imaginary parts of the second harmonic of the complex valued nano-FTIR spectrum of PS referenced to Si: $\bar{z}_2 = \mathfrak{R} z_2^{PS} = z_2^{PS}/z_2^{Si}$. Note that the former possesses dispersion-like features, while the later possess absorption-like features.

***Fourier Transformation $\mathfrak{F}$: Removal of Self-Homodyne Background and Generation of the Unnormalized Complex Valued Spectrum.*** As shown in Fig. 1 and described above, the

interferogram $I'_{n>1}(\xi)$ (the real part of the complex valued interferogram) is passed along to a computer for further processing where Fourier transformation (FT) and normalization occur. Thus, the next computation step is to conduct the relevant FT: $\mathfrak{F}_{\xi \to \nu} I'_{n>1}(\xi)$, where $\mathfrak{F}_{\xi \to \nu}$ is the operator for FT and for some function of $\xi$, say $p(\xi)$, the FT is here defined as

$$\mathfrak{F}_{\xi \to \nu} p(\xi) = \int_{-\infty}^{\infty} \{p(\xi_1) e^{i2\pi\nu\xi_1}\} d\xi_1. \tag{16}$$

Ultimately, this operation will (i) remove the self-homodyne background signal (terms proportional to expansion coefficients $V_n$ that contain interference/coupling between $E^{FF}$ and $E^{NF}$), (ii) change the independent variable from $\xi$ to wavenumber '$\nu$', and (iii) generate the unnormalized complex valued nano-FTIR spectrum. Because the FT operation on a constant is known to produce a Dirac delta distribution, and because the first term in the real part of complex interferogram (15) is independent of $\xi$, that term vanishes for non-zero wavenumbers; and so, the self-homodyne background vanishes. This leaves behind the FT of the second term, the heterodyne term: $\mathfrak{F}_{\xi \to \nu} I'_{n>1}(\xi) = \mathfrak{F}_{\xi \to \nu} u_n^{NF}(\xi) \cos(\theta_n(\xi))/2$. However, this expression can be simplified further because $u_n^{NF}(\xi)$ is inversely proportional to the cosine term, as shown in Equation (11), and thus $\mathfrak{F}_{\xi \to \nu} I'_{n>1}(\xi) = \mathfrak{F}_{\xi \to \nu} u_n^{NF}(\xi; \theta_n = 0)/2$. So,

$$z_{n>1}(\nu) \equiv \mathfrak{F}_{\xi \to \nu} I'_{n>1}(\xi) = \mathfrak{F}_{\xi \to \nu} \frac{u_n^{NF}(\xi; \theta_n = 0)}{2}, \tag{17}$$

where $z_{n>1}(\nu)$ is here defined as the raw, unnormalized, and complex valued nano-FTIR spectrum. Examples of the real and imaginary parts of the raw, unnormalized, and complex valued nano-FTIR spectrum (2nd harmonic) for Si and PS are plotted in Figures 5c and 5d respectively.

Now, with our knowledge of how the electric fields from the sample and reference arms relate to the expansion coefficients $u_n^{NF}$, we can move to compute the complex nano-FTIR spectrum in terms of these fields. Utilizing Equations (11), (16), and (17), and changing the integration order yields

$$z_{n>1}(\nu) = \frac{\eta a \operatorname{sinc}\left(\frac{T_d n \omega_c}{2}\right)}{2\pi c \mu_0} \int_{-\infty}^{\infty} [\tilde{E}_y^{NF*}(\omega - n\omega_c) + \tilde{E}_y^{NF*}(\omega + n\omega_c)] \tilde{E}_{RA,y}(\omega) \int_{-\infty}^{\infty} e^{i\xi_1\left(\frac{\omega}{c} + 2\pi\nu\right)} d\xi_1 \, d\omega,$$

which simplifies to

$$z_{n>1}(\nu) = \Gamma_n [\tilde{E}_y^{NF}(2\pi\nu c + n\omega_c) + \tilde{E}_y^{NF}(2\pi\nu c - n\omega_c)] \tilde{E}_{RA,y}^*(2\pi\nu c), \tag{18}$$

where $\Gamma_n = 2\eta a \text{sinc}(T_d n\omega_c/2)/(c\mu_0)$. The simplification process to Equation (18) included (i) the integral over $\xi_1$ reducing to a Dirac delta distribution which demanded $\omega \to -2\pi\nu c$, and (ii) the use of the reality condition that if $f(t)$ is real-valued, $\tilde{f}^*(-\omega) = \tilde{f}(\omega)$.

***Normalization $\mathfrak{N}$: Removal of Scaled Reference Arm Light and Arrival at the Normalized Complex Valued Spectrum.*** The last processing step is normalization. As can be seen in Equation (18), an unnormalized nano-FTIR spectrum is linearly proportional to $\Gamma_n \tilde{E}^*_{RA,y}(2\pi\nu c)$. To remove these, a normalization strategy is employed. Two nano-FTIR spectra are collected, one of the sample material of interest, and another of a reference material with constant-in-$\nu$ (or smoothly varying) dielectric properties. Then, a ratio of the two spectra is taken, and the $\Gamma_n \tilde{E}^*_{RA,y}(2\pi\nu c)$ terms in the numerator and denominator cancel as both reference field and prefactor $\Gamma_n$ do not change between measurements. Thus, in general, a normalized complex valued nano-FTIR spectrum, $\mathfrak{N} z_{n>1}(\nu) = \bar{z}_{n>1}(\nu)$, will take the form,

$$\bar{z}_{n>1}(\nu) \equiv \frac{z_{n>1}^{t\&sm}(\nu)}{z_{n>1}^{t\&rm}(\nu)} = \frac{\tilde{E}_y^{NF,t\&sm}(\omega_n^+(\nu)) + \tilde{E}_y^{NF,t\&sm}(\omega_n^-(\nu))}{\tilde{E}_y^{NF,t\&rm}(\omega_n^+(\nu)) + \tilde{E}_y^{NF,t\&rm}(\omega_n^-(\nu))}, \quad (19)$$

where superscripts "$t\&sm$" and "$t\&rm$" stand for "tip and sample material" and "tip and reference material," respectively, and $\omega_n^\pm(\nu) = 2\pi\nu c \pm n\omega_c$.

We note that Equation (19) can also be cast in terms of the FT of the product of $E_y^{NF}$ and $\cos(n\omega_c t)$:

$$\bar{z}_{n>1}(\nu) = \frac{\mathfrak{F}_{t\to\omega(\nu)}[E_y^{NF,t\&sm}(t)\cos(n\omega_c t)]}{\mathfrak{F}_{t\to\omega(\nu)}[E_y^{NF,t\&rm}(t)\cos(n\omega_c t)]}. \quad (20)$$

As mentioned earlier, a "good" reference material will be spectrally flat, typically having no absorption resonances within the wavenumber region of interest. Common reference materials are gold or silicon. So, as to provide a concrete example, if the material of interest was PS, the reference material was Si, and $n$ was two, a thorough way to refer to such a case would be "the second harmonic of the complex valued nano-FTIR spectrum of PS referenced to Si." The real and imaginary parts of this spectrum, $\bar{z}_2'(\nu)$ and $\bar{z}_2''(\nu)$, are plotted in Figure 5e and 5f respectively.

In principle, the right-hand side of either equation (19) or (20) can be used as a springboard for inputting one's favorite near-field scattering model. This would be done, of course, with the aim to establish relationships between the sample material's local properties, and the real and imaginary parts of $\bar{z}_{n>1}(\nu)$, which are measured. However, in their current forms, they lack explicit dependence on physical quantities that are associated with, or reside in, the tip-sample region. We address this in the following.

***Casting the Normalized Complex Valued Spectrum in Terms of Dipole Radiation from a Nanoscale Volume.*** Now that the detection and signal processing steps for scattering-type near-field infrared nanospectroscopy have been fully modeled, we aim to move toward interpretation. This requires casting $\tilde{E}_y^{NF}(\omega) = \tilde{E}_y^{NF}(\omega, x = 0)$, a quantity physically located at the detector surface, in terms of quantities within the tip-sample region. For this reason, we first define new cartesian and polar coordinate systems that more directly relate to the scattering problem, and whose origin is the center of the tip-sample region at minimum tip-sample separation, as shown in Fig. 6.

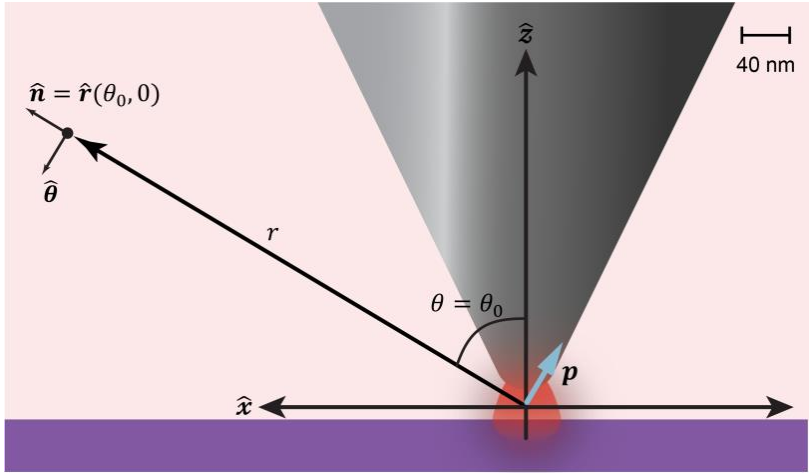

**Fig. 6** Schematic of the tip-sample region.

Because (i) the detector is physically located at a much larger distance ($R$) than the characteristic size ($\ell$) of the charges accelerating due to near-field excitations (they are confined to nanoscopic volumes in the tip-sample region), (ii) the wavelength of the incident IR light is also much larger then $\ell$, and (iii) $R \gg \lambda \gg \ell$, dipole radiation[51-53] from the tip-sample region is what constitutes $\tilde{E}_y^{NF}(\omega, x = 0)$. Thus, we draw (as a blue arrow in Fig. 6) an effective net dipole moment, $p$, of the whole tip-sample region as would be "seen" a far distance away. From symmetry, we infer $p$ should be confined to the $x/z$ plane because the incident polarization is primarily in the $x/z$ plane. The mathematical relationship between the coordinate systems can be expressed as

$$\tilde{E}_y^{NF}(\omega, x = 0) = \tilde{E}_\theta^{NF}(\omega, R, \theta_0, 0) g(\omega), \qquad (21)$$

where $R$ is the fixed distance between the tip-sample region and the detector, $\theta_0$ is the fixed angle between the polar axis ($\hat{z}$ in Fig. 6) and the fixed backscattering direction $\hat{n} = \hat{r}(\theta_0, 0)$ (see Fig. 6), and $g(\omega)$ is a correction factor which accounts for influences that the optics of the sample arm of the AMI may have on the backscattered light as it travels towards the detector. Furthermore, generalized dipole radiation fields are well-known in terms of the FTs in time.[52,53] For the case outlined above, this vector expression is $(-\omega^2 \mu_0 Exp[i\omega r/c](\tilde{p} \times \hat{n}) \times \hat{n}))/(4\pi R)$, it reduces to a vector expression having only a $\hat{\theta}$ component. In particular,

$$\tilde{E}_\theta^{NF}(\omega, R, \theta_0, 0) = \frac{-\omega^2 \mu_0 e^{\frac{i\omega R}{c}}}{4\pi R}(-\tilde{p}_x(\omega) \cos[\theta_0] + \tilde{p}_z(\omega) \sin[\theta_0]), \qquad (22)$$

where $\tilde{p}_x$ and $\tilde{p}_z$ are the vector components of $\widetilde{\boldsymbol{p}}$.

With equations (21) and (22) input into (19), and with reasonable simplifications outlined in the Supporting Information, the normalized complex valued spectrum becomes

$$\bar{z}_{n>1}(\nu) = \frac{\left(\frac{n\omega_c}{\omega}\right)^0 \Delta_+^{t\&sm} + \left(\frac{n\omega_c}{\omega}\right)^1 2\,\Delta_-^{t\&sm} + \left(\frac{n\omega_c}{\omega}\right)^2 \Delta_+^{t\&sm}}{\left(\frac{n\omega_c}{\omega}\right)^0 \Delta_+^{t\&rm} + \left(\frac{n\omega_c}{\omega}\right)^1 2\,\Delta_-^{t\&rm} + \left(\frac{n\omega_c}{\omega}\right)^2 \Delta_+^{t\&rm}}, \qquad (23)$$

where,

$$\Delta_\pm^j = \frac{\mathcal{D}^j(\omega + n\omega_c) \pm \mathcal{D}^j(\omega - n\omega_c)}{2n\omega_c},$$

and

$$\mathcal{D}^j(\omega) = e^{\frac{i\omega R}{c}} g(\omega)\left(-\tilde{p}_x^j(\omega)\,\cot[\theta_0] + \tilde{p}_z^j(\omega)\right).$$

In the above, the superscripts on $\mathcal{D}^j$ and $\Delta_\pm^j$ take one of the two already defined notations of $j = t\&sm$ or $t\&rm$ and we stop explicitly notating functional dependence of $\omega$ on $\nu$ (treating this as understood). With this, we have finally cast the normalized spectrum in terms of clear-cut (though yet undetermined) physical quantities associated with the tip-sample region, $\tilde{p}_x$ and $\tilde{p}_z$, which can be modeled as one sees fit. As currently expressed, each of the three terms in the numerator (and denominator) of the normalized complex valued spectrum are scaled by a small dimensionless multiplicative prefactor, raised to various powers (say of $q$): $(n\omega_c/\omega_{IR})^q$. For the system at hand, $10^{-9} \lesssim (n\omega_c/\omega) \lesssim 10^{-8}$. Therefore, the power $q$ of the prefactor defines each terms' relative degree of smallness: zeroth order in smallness ($q = 0$), first order in smallness ($q = 1$), and second order in smallness ($q = 2$). We additionally mention that each of the $\Delta_\pm^j$ terms in (23), with a judicious use of Taylor series expansions, can be approximated as a series that contain various orders of derivatives involving the dipole moments. That said, such pursuits toward ultimate numerical precision are beyond the scope of this present work, but may be relevant in some physical situations.

Equation (23) should be considered a robust springboard/starting point for quantitative modeling of normalized spectra empirically collected. We emphasize that at this stage we have made no assumptions of any kind as to functional forms of the dipole terms in $\mathcal{D}$, they are strictly general expressions, and can accommodate any number of models. In order to extract specific information about the sample material's local properties, "all" that needs to be done is the following. First, choose by order of smallness which terms in (23) to keep. Second, for the terms that remain, generate and input models for the quantities on the right-hand side of (23). These models will be complex-valued as expressions on the right-hand side of (23) depend on frequency-dependent FTs and complex exponentials. Third, equate the real and imaginary parts of the model to the corresponding empirical data on the left-hand side of (23) (symbolized by $\bar{z}_{n>1}(\nu)$). Lastly, choose an appropriate method, for each frequency value, to solve the system of two equations for two unknown model parameters of interest; these parameters are usually the real and imaginary parts of the sample's relative dielectric function, $\tilde{\epsilon}_r^{sm} = \tilde{\epsilon}_r^{sm'} + i\tilde{\epsilon}_r^{sm''}$.

***Considering Detected Radiation as Originating Primarily from a Net Dipole Radiator Oriented Parallel to the Probe Axis.*** Usually, contributions from dipoles oriented parallel to the sample surface ($\tilde{p}_x$) are excluded in literature analyses, as their contributions to the detected radiation, though non-zero, can be small in comparison to those originating from $\tilde{p}_z$.[54] In light of this, and in keeping with our move toward gaining a qualitative interpretation of $\bar{z}_{n>1}$, we drop dipoles with $x$ dependence from consideration. In such a case, $\boldsymbol{p}$ in Fig. 6 will align with the polar axis $\hat{z}$, and $\tilde{p}_x \to 0$ in $\mathcal{D}^j(\omega)$ terms in (23). Additionally, we will consider two more simplifications. First, we point out that $n\omega_c R/c \sim 10^{-2}$ for the system at hand, so $\exp[i\omega_n^\pm R/c] \approx \exp[i\omega R/c]$ and the complex exponentials in $\mathcal{D}^j(\omega)$ terms in (23) cancel. Secondly, we consider that optical-component influences on the backscattered light only weakly change with $\omega$, and so $g(\omega_n^\pm) \to g(\omega)$. Application of these lead to (23) simplifying to

$$\bar{z}_{n>1}(\nu) = \frac{\left(\frac{n\omega_c}{\omega}\right)^0 \delta_+^{t\&sm} + \left(\frac{n\omega_c}{\omega}\right)^1 2\delta_-^{t\&sm} + \left(\frac{n\omega_c}{\omega}\right)^2 \delta_+^{t\&sm}}{\left(\frac{n\omega_c}{\omega}\right)^0 \delta_+^{t\&rm} + \left(\frac{n\omega_c}{\omega}\right)^1 2\delta_-^{t\&rm} + \left(\frac{n\omega_c}{\omega}\right)^2 \delta_+^{t\&rm}} \tag{24}$$

with,

$$\delta_\pm^j = \frac{\tilde{p}_z^j(\omega + n\omega_c) \pm \tilde{p}_z^j(\omega - n\omega_c)}{2n\omega_c}.$$

Equation (24) is as fully applicable as (23), so long as only $\tilde{p}_z$ need be considered, and the aforementioned approximation of $g(\omega_n^\pm) \to g(\omega)$, is appropriate.

Next, we point out that the FT of the net dipole of the tip-sample region, $\tilde{\boldsymbol{p}} = \tilde{p}_z^{t\&sm}\hat{\boldsymbol{z}}$, can generally be found with a volume integral of the $z$-component of the polarization density: $\tilde{p}_z^{t\&sm} = \int_V \tilde{P}_z\, dV$ (see Fig. 7a). We note that a rigid definition of the exact integration volume is not needed for our purposes here, though, of course, strict boundary definitions would be required for rigorous modeling of the scattering. We here sketch a qualitative representative outline of a boundary for such a volume, and the related net dipole along the $z$ axis, $\boldsymbol{p}$ in Fig. 7a.

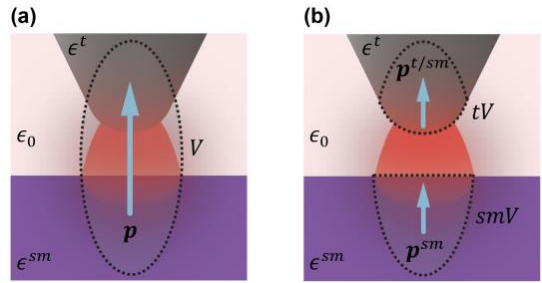

**Fig. 7.** Equivalent schematics/representations of the tip-sample region as perceived by an observer far away from the accelerating charges. Dielectric permittivity is appropriately labeled. (a) The net dipole moment induced within the volume $V$ by the enclosed charge distributions being subjugated to near-field excitations. (b) Two dipole moments induced within distinct volumes, $tV$ and $smV$, arising from the enclosed charge distributions being subjugated to near-field excitations (and dipole-dipole coupling).

Without loss of generality, the single integral over $V$ can be replaced by three volume integrals. One for each material: $\tilde{p}_z^{t\&sm} = \int_V \tilde{P}_z \, dV = \int_{tV} \tilde{P}_z \, dV + \int_{air} \tilde{P}_z \, dV + \int_{smV} \tilde{P}_z \, dV$, where $tV$ stands for the tip volume, air stands for the air gap between probe tip and sample, and $smV$ stands for sample material volume. Because the air in the tip-sample gap doesn't polarize, that integral vanishes, leaving $\tilde{p}_z^{t\&sm} = \int_{tV} \tilde{P}_z \, dV + \int_{smV} \tilde{P}_z \, dV = \tilde{p}_z^{t/sm} + \tilde{p}_z^{sm}$, and each remaining integral result, respectively, constitutes $\widetilde{\boldsymbol{p}}^{t/sm} = \tilde{p}_z^{t/sm}\hat{\boldsymbol{z}}$ and $\widetilde{\boldsymbol{p}}^{sm} = \tilde{p}_z^{sm}\hat{\boldsymbol{z}}$ (as shown in Fig. 7b). Thus, $\widetilde{\boldsymbol{p}} = \widetilde{\boldsymbol{p}}^{t/sm} + \widetilde{\boldsymbol{p}}^{sm}$, and $\tilde{p}_z^{t\&sm} = \tilde{p}_z^{t/sm} + \tilde{p}_z^{sm}$. Being able to separate out these two contributors to the net dipole moment is important for a holistic understanding, but also for developing detailed models which aim to account for dipole-dipole coupling though the electric field. Since the detection and signal processing theory at this stage is now complete, and with Equation (23) describing the normalized spectrum allowing for dipoles oriented both in and out of the sample surface's plane, and Equation (24) specializing to the case of net dipole/dipoles oriented out of the sample surface's plane, and parallel to the probe axis, we move toward the last goal of this work: deriving a basic, yet practically useful, interpretation of $\bar{z}_{n>1}$.

***Approximating the Complex Valued nano-FTIR Spectrum of Local Molecular Vibrations in a Nonmetal Sample Normalized with a Nonmetal Reference.*** Perhaps the most common application of nano-FTIR is to conduct vibrational spectroscopy of insulating or semi-conducting materials (nonmetals). This common class of measurements, aimed at identifying chemistry and other physicochemical properties, benefits from being able to compare nano-FTIR spectra with other conventional FTIR measurements for data interpretation. This process hinges on having a basic understanding of what $\bar{z}_{n>1}(\nu)$ physically represents. In this section we provide a semi-quantitative description of what $\bar{z}_{n>1}(\nu)$ most closely physically represents in the aforementioned case of bulk nonmetal samples and references. A simple approximation to Equation (24) is used to do so.

We treat the net dipole as being oriented parallel to the probe axis ($z$ axis in Fig. 6), as was the case in the last section. Furthermore, we consider an ideal case in which the sample material, reference material, and probe tip are optically isotropic; that is, we assume the dielectric function of all materials only depends on scalar frequency ($\nu$ or $\omega$). With this in mind, we now move to approximate (24) by (i) only keeping terms to zeroth order in smallness ($q = 0$), (ii) allowing $\tilde{p}_z^j(\omega + n\omega_c) + \tilde{p}_z^j(\omega - n\omega_c) \cong 2\tilde{p}_z^j(\omega)$, and (iii) replacing the single net dipole with two distinct dipoles (as described in the previous section and pictorially represented in Fig. 7b); which yields

$$\bar{z}_{n>1} \cong \frac{\tilde{p}_z^{t/sm} + \tilde{p}_z^{sm}}{\tilde{p}_z^{t/rm} + \tilde{p}_z^{rm}}. \tag{25}$$

The superscript notation $t/sm$ serves to indicate "tip adjacent to sample material" and similarly, $t/rm$ indicates "tip adjacent to reference material." Equation (26) shows that the normalized complex valued nano-FTIR spectrum is dominated by a ratio of the FT of the net dipole moment of the tip-sample region to that of the FT of the net dipole moment of the tip-reference material region.

Next, we use a general relation for the FT of an arbitrary dipole moment (derived in the Supporting Information): $\tilde{p}_z = \epsilon_0(\tilde{\epsilon}_r - 1)\xi_V$, where $\xi_V = \int_V \tilde{E}_z \, dV$. This yields

$$\bar{z}_{n>1} \cong \frac{\int_{tV} \tilde{E}_{z,in}^{t/sm} \, dV + \left(\frac{\tilde{\epsilon}_r^{sm} - 1}{\tilde{\epsilon}_r^{t} - 1}\right)\xi_V^{sm}}{\int_{tV} \tilde{E}_{z,in}^{t/rm} \, dV + \left(\frac{\tilde{\epsilon}_r^{rm} - 1}{\tilde{\epsilon}_r^{t} - 1}\right)\xi_V^{rm}}; \tag{26}$$

however, since the probe tip is considered a very good conductor, and if we treat cases where the sample and reference materials are both non-metallic, then the ratio in front of the second term in both the numerator and denominator is very small. Thus, we only need treat the first term in the numerator and denominator in (26): $\bar{z}_{n>1} \cong \xi_{tV}^{t/sm}/\xi_{tV}^{t/rm}$ (this is a manifestation of the fact that a metallic tip will generally backscatter much more light than a non-metallic sample, and so radiation originating from the tip's dipole dominates the detected signal). Still, a more advantageous form can be found. To do so, we treat the tip end as a sphere being polarized by both the incident IR light and the near electric field arising from the induced dipole moment in the sample. Additionally, the volume integral of the $z$ component of the FT of the electric field inside the sample and reference materials are approximated as smoothing varying in energy so that $\xi_V^{sm}/\xi_V^{rm} \approx$ constant. While all the details are clearly outlined in the Supporting Information, the procedure results in

$$\bar{z}_{n>1} \cong \frac{(\tilde{\epsilon}_r^{sm} - 1)\xi_V^{sm}}{(\tilde{\epsilon}_r^{rm} - 1)\xi_V^{rm}} \approx \frac{\tilde{\epsilon}_r^{sm} - 1}{\tilde{\epsilon}_r^{rm} - 1}. \tag{27}$$

Equation (27) is surprisingly simple, and naturally lends itself to an algebraically straightforward way of explaining why $\bar{z}'_{n>1}$ closely relates to dispersion, and $\bar{z}''_{n>1}$ closely relates to absorption. Furthermore, as we will show, algebraic inversion of (27), $\bar{z}_{n>1}(\tilde{\epsilon}_r^{sm}, \tilde{\epsilon}_r^{rm}) \to \tilde{\epsilon}_r^{sm}(\bar{z}_{n>1}, \tilde{\epsilon}_r^{rm})$, provides a simple pathway to transform nano-FTIR data into a model extinction coeffect that closely matches empirical results measured via attenuated total reflection Fourier transform infrared spectroscopy (ATR-FTIR).

Assuming that the reference material being used is non-metallic, spectrally flat in the IR wavenumber region of interest, and also realizes negligible absorption in the IR wavenumber region of interest, the real (imaginary) part of the reference material's dielectric function remains constant-in-energy, $d\tilde{\epsilon}_r^{rm'}/dv \approx 0$ (vanishes, $\tilde{\epsilon}_r^{rm''} \approx 0$). This is the case for the commonly used reference material of crystalline Si (cSi): $\tilde{\epsilon}_r^{rm} = \tilde{\epsilon}_r^{cSi} \cong 11.7 + i\, 8.48 * 10^{-4} \approx 11.7$.[55] So, assuming this kind/class of reference material with vanishingly small $\tilde{\epsilon}_r^{rm''}$, the real and imaginary parts of $\bar{z}_{n>1}$ are, approximately,

$$\bar{z}'_{n>1} \approx \frac{\tilde{\epsilon}_r^{sm'} - 1}{\tilde{\epsilon}_r^{rm'} - 1} \quad , \quad \bar{z}''_{n>1} \approx \frac{\tilde{\epsilon}_r^{sm''}}{\tilde{\epsilon}_r^{rm'} - 1}. \tag{28}$$

Apparently the real and imaginary parts of $\bar{z}_{n>1}$ directly relate to $\tilde{\epsilon}_r^{sm}$, the sample material's relative complex dielectric permittivity (i.e. dielectric function): $\bar{z}'_{n>1} \propto \tilde{\epsilon}_r^{sm'} - 1$ and $\bar{z}''_{n>1} \propto \tilde{\epsilon}_r^{sm''}$. Because the real part of the dielectric function is closely related to dispersion via the refractive index, $n$,[52,56] equation (28) qualitatively explains why the real part of $\bar{z}_{n>1}$ has a similar line shape and why it is common to report the real part of $\bar{z}_{n>1}$ as quantities relating to dispersion. Similarly, because the imaginary part of the dielectric function is closely related to absorption via the extinction coefficient, $\kappa$,[52,56] equation (28) qualitatively explains why the imaginary part of $\bar{z}_{n>1}$ has a similar line shape and why it is common to report the imaginary part of $\bar{z}_{n>1}$ as relating to absorption. These relationships are easily observed in Fig. 5e and f, and become even more acute in the weak oscillator limit ($n \gg \kappa$), where $n \approx \sqrt{\epsilon'_r}$ and $\kappa \approx \epsilon''_r/(2\sqrt{\epsilon'_r})$[56,57]; as others have also shown by different means.[11,26] However, even in the weak oscillator limit, $\kappa$ still depends on both the real and imaginary parts of the dielectric function, and since far-field transmission-type FTIR absorbance spectra measure $\kappa$,[57,58] it stands to reason that $\bar{z}''_{n>1} \propto \tilde{\epsilon}_r^{sm''}$ alone may not be the best near-field quantity to report as absorption, and compare with other FTIR spectral databases.

The level of connection between $\bar{z}''_2$ and $\kappa$ for a PS sample is compared in Fig. 8a by plotting $\bar{z}''_2$ (referenced with respect to cSi) in red alongside ATR-FTIR absorbance data of the same sample in black (spectra are normalized to unity). While the $\bar{z}''_2$ peak at 692 cm$^{-1}$ is red-shifted ~2.4 cm$^{-1}$ in comparison to ATR-FTIR absorbance, the similarity is still quite striking, and gives credence to the use of $\bar{z}''_2$ as a proxy for absorption, as expected. However, a better agreement can be found with a simple algebraic manipulation of the near-field data. By solving the two equations in (28) for the two unknowns, we can establish functions for real and imaginary parts of the sample's dielectric function in terms of the real and imaginary parts of $\bar{z}_n$: $\tilde{\epsilon}_r^{sm'}(\bar{z}'_n, \bar{z}''_n)$ and $\tilde{\epsilon}_r^{sm''}(\bar{z}'_n, \bar{z}''_n)$. These can be used to compute an expression for the model extinction coefficient (or refractive index) in terms of $\bar{z}'_n$ and $\bar{z}''_n$; this is because the extinction coefficient is generally a function of the real and imaginary parts of the dielectric function:[56] $\kappa(\tilde{\epsilon}_r^{sm'}, \tilde{\epsilon}_r^{sm''}) \to \kappa(\bar{z}'_n, \bar{z}''_n)$. The normalized result, when considering a cSi reference material as given above, is

$$\bar{\kappa}_{n>1} \cong b\left[\sqrt{114.5\,\bar{z}''_n{}^2 + (1 + 10.7\,\bar{z}'_n)^2} - (1 + 10.7\,\bar{z}'_n)\right]^{1/2}, \qquad (29)$$

where $b$ serves as the constant scaling/normalization parameter; and we note that the 10.7 prefactor to $\bar{z}'_n$ in (29) will be different for different reference materials. As can be seen in equation (29), $\bar{\kappa}_{n>1}$ depends on both the real and imaginary parts of $\bar{z}_{n>1}$. Inputting the nano-FTIR data for PS (Fig. 5e and f) into (29) yields a result that is plotted in blue in Fig. 8a. The red-shift relative to the ATR-FTIR data has essentially vanished (~0.1 cm$^{-1}$).

The validity and utility of (29) was further tested by conducting ATR-FTIR and nano-FTIR of a Kapton sample. The results are plotted in Fig. 8b. As in the PS case above, $\bar{\kappa}_2$ is a good approximation of the ATR-FTIR absorbance spectrum which measures $\kappa$, and while perhaps more challenging to see on the broader scale, $\bar{\kappa}_2$ is again a better estimate of ATR-FTIR peak centers than $\bar{z}''_2$ alone. In particular, the average $\bar{\kappa}_2$ peak center is 17% closer to ATR-FTIR peak centers than $\bar{z}_2''$ peak centers (based on an analysis of eight peaks fitted with Lorenz models – see Supporting Information Fig. S1). The increased peak center accuracy that (29) provides may be found useful when trying to assign spectral features of nano-FTIR spectra to specific molecular oscillations. As one final note, we stress that this approximation is most applicable for nano-FTIR of weak and local molecular oscillations in nonmetal materials (like polymers) referenced to a spectrally flat nonmetal (like cSi).

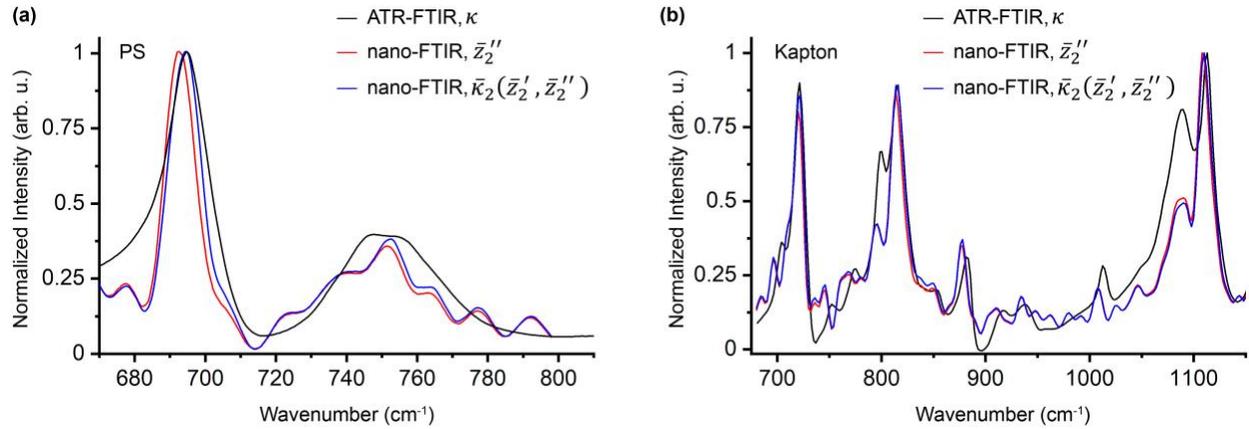

**Fig. 8.** ATR-FTIR, nano-FTIR, and $\kappa_2(\bar{z}_2', \bar{z}_2'')$ extracted from the real and imaginary parts of the nano-FTIR spectrum for (a) polystyrene and (b) Kapton.

### 3. Conclusion

At an increasing rate, nano-FTIR is being utilized to conduct nanoscale infrared characterization over an impressive breadth of fields, including meteoritics, chemistry, physics, energy storage, and biology. The broad potential applicability of the method is hard to overstate. In this work, we have provided holistic and quantitative derivations of how nano-FTIR spectra are realized, presenting a new and self-contained work that benefits both beginning and seasoned practitioners. We introduced the technique and rigorously stepped through the detection and signal processing steps. Along the way, common questions are naturally answered, and connections to past works are highlighted. A general relation for normalized complex valued nano-FTIR spectra, in terms of FTs of general dipole moments in the tip-sample region oriented both in and out of the sample surface plane, was provided. This formalism can be used as a spring board for additional modeling endeavors. Moreover, we developed a new, surprisingly simple yet insightful model, valid (at least) in the weak oscillator limit when using a cSi reference. It rationalizes why the real and imaginary parts of complex valued nano-FTIR spectra relate to dispersion and absorption respectively, and facilitates an approximation to the samples' local extinction coefficient that more closely matches ATR-FTIR data than the imaginary part alone. The extinction coefficient model herein takes as inputs bot the real and imaginary parts of normalized complex valued nano-FTIR spectra, and can be applied straightforwardly with algebra, without need for computations.

# SUPPORTING INFORMATION

Supporting Information detailing mathematical derivations related to this work, and called out above in the main text, is available in the supporting information below.

# ACKNOWLEDGMENTS

We kindly acknowledge sources that financially supported this work. Funding to support this work was provided to J.M.L., H.A.B., and R.K. by the Energy & Biosciences Institute through the EBI-Shell program. Additionally, funding to support this work was also provided to J.M.L and R.K. from the Assistant Secretary for Energy Efficiency and Renewable Energy, Vehicle Technologies Office, under the Advanced Battery Materials Research (BMR) Program, of the U.S. Department of Energy under Contract No. DE-AC02-05CH11231.

This research also used resources of the Advanced Light Source, a U.S. DOE Office of Science User Facility under contract no. DE-AC02-05CH11231. In particular, Beamlines 2.4 and 5.4 were utilized.

# CONFLICT OF INTEREST

The authors declare no conflict of interest.

# DATA AVAILABILITY STATEMENT

All data to support the findings of this study are provided either in the main text or the Supporting Information.

# REFERENCES

~References~

1. K. Nakamoto. *Infrared and Raman Spectra of Inorganic and Coordination Compounds*. (John Wiley & Sons, 2009).
2. T. Hasegawa. *Quantitative Infrared Spectroscopy for Understanding of a Condensed Matter*. (2017).
3. J. M. Thompson. *Infrared Spectroscopy*. (Pan Stanford Publishing, 2018).
4. H. J. Humecki. *Practical guide to infrared microspectroscopy*. (CRC Press, 1995).

5. J. E. Katon. Infrared microspectroscopy. A review of fundamentals and applications. *Micron* **27**, 303-314 (1996).
6. G. L. Carr. Resolution limits for infrared microspectroscopy explored with synchrotron radiation. *Rev. Sci. Instrum.* **72** (2001).
7. B. C. Smith. *Fundamentals of Fourier Transform Infrared Spectroscopy*. (CRC Press, 2011).
8. P. M. Ajayan. The nano-revolution spawned by carbon. *Nature* **575**, 49-50 (2019).
9. X. Chen, D. Hu, R. Mescall, G. You, D. N. Basov, Q. Dai & M. Liu. Modern Scattering-Type Scanning Near-Field Optical Microscopy for Advanced Material Research. *Adv. Mater.* **31**, 1804774 (2019).
10. J. M. Atkin, S. Berweger, A. C. Jones & M. B. Raschke. Nano-optical imaging and spectroscopy of order, phases, and domains in complex solids. *Advances in Physics* **61**, 745-842 (2012).
11. F. Huth, A. Govyadinov, S. Amarie, W. Nuansing, F. Keilmann & R. Hillenbrand. Nano-FTIR absorption spectroscopy of molecular fingerprints at 20 nm spatial resolution. *Nano Lett.* **12**, 3973-3978 (2012).
12. H. A. Bechtel, E. A. Muller, R. L. Olmon, M. C. Martin & M. B. Raschke. Ultrabroadband infrared nanospectroscopic imaging. *Proc Natl Acad Sci U S A* **111**, 7191-7196 (2014).
13. A. Centrone. Infrared Imaging and Spectroscopy Beyond the Diffraction Limit. *Annu Rev Anal Chem (Palo Alto Calif)* **8**, 101-126 (2015).
14. F. Huth. Nano-FTIR-Nanoscale Infrared Near-Field Spectroscopy. *Euskal Herriko Unibertsitatea-Universidad del Pais Vasco* (2015).
15. E. A. Muller, B. Pollard & M. B. Raschke. Infrared Chemical Nano-Imaging: Accessing Structure, Coupling, and Dynamics on Molecular Length Scales. *J. Phys. Chem. Lett.* **6**, 1275-1284 (2015).
16. O. Khatib, H. A. Bechtel, M. C. Martin, M. B. Raschke & G. L. Carr. Far Infrared Synchrotron Near-Field Nanoimaging and Nanospectroscopy. *ACS Photonics* **5**, 2773-2779 (2018).
17. H. A. Bechtel, S. C. Johnson, O. Khatib, E. A. Muller & M. B. Raschke. Synchrotron infrared nano-spectroscopy and -imaging. *Surf. Sci. Rep.* **75** (2020).
18. A. F. Moslein, M. Gutierrez, B. Cohen & J. C. Tan. Near-Field Infrared Nanospectroscopy Reveals Guest Confinement in Metal-Organic Framework Single Crystals. *Nano Lett.* **20**, 7446-7454 (2020).
19. Y. H. Lu, J. M. Larson, A. Baskin, X. Zhao, P. D. Ashby, D. Prendergast, H. A. Bechtel, R. Kostecki & M. Salmeron. Infrared Nanospectroscopy at the Graphene-Electrolyte Interface. *Nano Lett.* **19**, 5388-5393 (2019).
20. C. Y. Wu, W. J. Wolf, Y. Levartovsky, H. A. Bechtel, M. C. Martin, F. D. Toste & E. Gross. High-spatial-resolution mapping of catalytic reactions on single particles. *Nature* **541**, 511-515 (2017).
21. Z. Yao, X. Chen, L. Wehmeier, S. Xu, Y. Shao, Z. Zeng, F. Liu, A. S. McLeod, S. N. Gilbert Corder, M. Tsuneto, W. Shi, Z. Wang, W. Zheng, H. A. Bechtel, G. L. Carr, M. C. Martin, A. Zettl, D. N. Basov, X. Chen, L. M. Eng, S. C. Kehr & M. Liu. Probing subwavelength in-plane anisotropy with antenna-assisted infrared nano-spectroscopy. *Nat. Commun.* **12**, 2649 (2021).

22. S. Dai, Z. Fei, Q. Ma, A. S. Rodin, M. Wagner, A. S. McLeod, M. K. Liu, W. Gannett, W. Regan, K. Watanabe, T. Taniguchi, M. Thiemens, G. Dominguez, A. H. Castro Neto, A. Zettl, F. Keilmann, P. Jarillo-Herrero, M. M. Fogler & D. N. Basov. Tunable phonon polaritons in atomically thin van der Waals crystals of boron nitride. *Science* **343**, 1125-1129 (2014).
23. A. Fali, S. T. White, T. G. Folland, M. He, N. A. Aghamiri, S. Liu, J. H. Edgar, J. D. Caldwell, R. F. Haglund & Y. Abate. Refractive Index-Based Control of Hyperbolic Phonon-Polariton Propagation. *Nano Lett.* **19**, 7725-7734 (2019).
24. M. K. Liu, M. Wagner, E. Abreu, S. Kittiwatanakul, A. McLeod, Z. Fei, M. Goldflam, S. Dai, M. M. Fogler, J. Lu, S. A. Wolf, R. D. Averitt & D. N. Basov. Anisotropic electronic state via spontaneous phase separation in strained vanadium dioxide films. *Phys. Rev. Lett.* **111**, 096602 (2013).
25. B. Lyu, H. Li, L. Jiang, W. Shan, C. Hu, A. Deng, Z. Ying, L. Wang, Y. Zhang, H. A. Bechtel, M. C. Martin, T. Taniguchi, K. Watanabe, W. Luo, F. Wang & Z. Shi. Phonon Polariton-assisted Infrared Nanoimaging of Local Strain in Hexagonal Boron Nitride. *Nano Lett.* **19**, 1982-1989 (2019).
26. A. A. Govyadinov, I. Amenabar, F. Huth, P. S. Carney & R. Hillenbrand. Quantitative Measurement of Local Infrared Absorption and Dielectric Function with Tip-Enhanced Near-Field Microscopy. *J. Phys. Chem. Lett.* **4**, 1526-1531 (2013).
27. A. Fali, S. Gamage, M. Howard, T. G. Folland, N. A. Mahadik, T. Tiwald, K. Bolotin, J. D. Caldwell & Y. Abate. Nanoscale Spectroscopy of Dielectric Properties of Mica. *ACS Photonics* **8**, 175-181 (2020).
28. M. Yesiltas, T. D. Glotch & B. Sava. Nano-FTIR spectroscopic identification of prebiotic carbonyl compounds in Dominion Range 08006 carbonaceous chondrite. *Sci. Rep.* **11**, 11656 (2021).
29. M. Schnell, M. Goikoetxea, I. Amenabar, P. S. Carney & R. Hillenbrand. Rapid Infrared Spectroscopic Nanoimaging with nano-FTIR Holography. *ACS Photonics* **7**, 2878-2885 (2020).
30. J. Qian, Y. Luan, M. Kim, K.-M. Ho, Y. Shi, C.-Z. Wang, Y. Li & Z. Fei. Nonequilibrium phonon tuning and mapping in few-layer graphene with infrared nanoscopy. *Physical Review B* **103** (2021).
31. I. T. Lucas, A. S. McLeod, J. S. Syzdek, D. S. Middlemiss, C. P. Grey, D. N. Basov & R. Kostecki. IR near-field spectroscopy and imaging of single Li(x)FePO4 microcrystals. *Nano Lett.* **15**, 1-7 (2015).
32. X. He, J. M. Larson, H. A. Bechtel & R. Kostecki. In situ infrared nanospectroscopy of the local processes at the Li/polymer electrolyte interface. *Nat. Commun.* **13**, 1398 (2022).
33. I. Amenabar, S. Poly, W. Nuansing, E. H. Hubrich, A. A. Govyadinov, F. Huth, R. Krutokhvostov, L. Zhang, M. Knez, J. Heberle, A. M. Bittner & R. Hillenbrand. Structural analysis and mapping of individual protein complexes by infrared nanospectroscopy. *Nat. Commun.* **4**, 2890 (2013).
34. X. Zhao, D. Li, Y. H. Lu, B. Rad, C. Yan, H. A. Bechtel, P. D. Ashby & M. B. Salmeron. In vitro investigation of protein assembly by combined microscopy and infrared spectroscopy at the nanometer scale. *Proc Natl Acad Sci U S A* **119**, e2200019119 (2022).

35. O. Khatib, J. D. Wood, A. S. McLeod, M. D. Goldflam, M. Wagner, G. L. Damhorst, J. C. Koepke, G. P. Doidge, A. Rangarajan, R. Bashir, E. Pop, J. W. Lyding, M. H. Thiemens, F. Keilmann & D. N. Basov. Graphene-Based Platform for Infrared Near-Field Nanospectroscopy of Water and Biological Materials in an Aqueous Environment. *ACS Nano* **9**, 7968-7975 (2015).
36. L. M. Meireles, I. D. Barcelos, G. A. Ferrari, A. N. P. A. A. de, R. O. Freitas & R. G. Lacerda. Synchrotron infrared nanospectroscopy on a graphene chip. *Lab Chip* **19**, 3678-3684 (2019).
37. K. J. Kaltenecker, T. Golz, E. Bau & F. Keilmann. Infrared-spectroscopic, dynamic near-field microscopy of living cells and nanoparticles in water. *Sci. Rep.* **11**, 21860 (2021).
38. B. O'Callahan, O. Qafoku, V. Balema, O. A. Negrete, A. Passian, M. H. Engelhard & K. M. Waters. Atomic Force Microscopy and Infrared Nanospectroscopy of COVID-19 Spike Protein for the Quantification of Adhesion to Common Surfaces. *Langmuir* **37**, 12089-12097 (2021).
39. N. Behr & M. B. Raschke. Optical Antenna Properties of Scanning Probe Tips: Plasmonic Light Scattering, Tip−Sample Coupling, and Near-Field Enhancement. *J. Phys. Chem. C* **112**, 3766-3773 (2008).
40. A. A. Govyadinov, G. Y. Panasyuk & J. C. Schotland. Phaseless three-dimensional optical nanoimaging. *Phys. Rev. Lett.* **103**, 213901 (2009).
41. A. S. McLeod, P. Kelly, M. D. Goldflam, Z. Gainsforth, A. J. Westphal, G. Dominguez, M. H. Thiemens, M. M. Fogler & D. N. Basov. Model for quantitative tip-enhanced spectroscopy and the extraction of nanoscale-resolved optical constants. *Physical Review B* **90** (2014).
42. P. McArdle, D. J. Lahneman, A. Biswas, F. Keilmann & M. M. Qazilbash. Near-field infrared nanospectroscopy of surface phonon-polariton resonances. *Physical Review Research* **2** (2020).
43. F. Mooshammer, M. A. Huber, F. Sandner, M. Plankl, M. Zizlsperger & R. Huber. Quantifying Nanoscale Electromagnetic Fields in Near-Field Microscopy by Fourier Demodulation Analysis. *ACS Photonics* **7**, 344-351 (2020).
44. I. Rajapaksa, K. Uenal & H. K. Wickramasinghe. Image force microscopy of molecular resonance: A microscope principle. *Appl. Phys. Lett.* **97** (2010).
45. F. S. Ruggeri, G. Longo, S. Faggiano, E. Lipiec, A. Pastore & G. Dietler. Infrared nanospectroscopy characterization of oligomeric and fibrillar aggregates during amyloid formation. *Nat. Commun.* **6**, 7831 (2015).
46. P. M. Donaldson, C. S. Kelley, M. D. Frogley, J. Filik, K. Wehbe & G. Cinque. Broadband near-field infrared spectromicroscopy using photothermal probes and synchrotron radiation. *Opt. Express* **24**, 1852-1864 (2016).
47. K. L. A. Chan, I. Lekkas, M. D. Frogley, G. Cinque, A. Altharawi, G. Bello & L. A. Dailey. Synchrotron Photothermal Infrared Nanospectroscopy of Drug-Induced Phospholipidosis in Macrophages. *Anal. Chem.* **92**, 8097-8107 (2020).
48. B. Voigtländer. *SCANNING PROBE MICROSCOPY*. (Springer, 2016).
49. T. J. Parker. Dispersive Fourier transform spectroscopy. *Contemporary Physics* **31**, 335-353 (1990).
50. D. J. Lahneman, T. J. Huffman, P. Xu, S. L. Wang, T. Grogan & M. M. Qazilbash. Broadband near-field infrared spectroscopy with a high temperature plasma light source. *Opt. Express* **25**, 20421-20430 (2017).

51. L. D. Landau. *The classical theory of fields*. Vol. 2 (Elsevier, 2013).
52. J. D. Jackson. *Classical Electrodynamics*. 3rd edn, (Wiley, 1998).
53. W. K. Panofsky & M. Phillips. *Classical electricity and magnetism*. (Courier Corporation, 2005).
54. N. Ocelic. *Quantitative near-field phonon-polariton spectroscopy*, Technische Universität München, (2007).
55. E. D. Palik. *Handbook of optical constants of solids*. Vol. 3 (Academic press, 1998).
56. M. Fox. *Optical Properties of Solids*. Second edn, (Oxford University Press, 2010).
57. S. Mastel, A. A. Govyadinov, T. V. A. G. de Oliveira, I. Amenabar & R. Hillenbrand. Nanoscale-resolved chemical identification of thin organic films using infrared near-field spectroscopy and standard Fourier transform infrared references. *Appl. Phys. Lett.* **106** (2015).
58. P. R. Griffiths & J. A. de Haseth. *Fourier Transform Infrared Spectrometry*. (2007).

# Detection and Signal Processing for Near-Field Nanoscale Fourier Transform Infrared Spectroscopy

*Jonathan M. Larson [‡] (ORCID: 0000-0002-5389-0794), Hans A. Bechtel [⊥,*] (ORCID: 0000-0002-7606-9333), and Robert Kostecki [‡,*] (ORCID: 0000-0002-4014-8232)*

[‡] Energy Storage & Distributed Resources Division, Lawrence Berkeley National Laboratory, Berkeley, California, 94720, United States

[⊥] Advanced Light Source, Lawrence Berkeley National Laboratory, Berkeley, California, 94720, United States

*Corresponding Authors: Hans A. Bechtel (habechtel@lbl.gov) and Robert Kostecki (R_Kostecki@lbl.gov)

## Mathematical Treatment of the Third Integral in Main Text Equation (6)

As mentioned in the main text, the third integral in equation (6) becomes,

$$\frac{\eta a}{2\pi c \mu_0} \int_{-\infty}^{\infty} \int_{-\infty}^{\infty} \tilde{E}_{R,y}(\omega_5) \tilde{E}_{R,y}(\omega_6) e^{i\xi\left(\frac{\omega_5}{c}+\frac{\omega_6}{c}\right)} e^{-it\Omega_{56}} \text{sinc}\left(\frac{T_d \Omega_{56}}{2}\right) d\omega_5 d\omega_6,$$

which can be reduced further. This is done by taking a moment to consider the sinc function, as well as the possible values for $\Omega_{56} = \omega_5 + \omega_6$ in which the amplitudes ($\tilde{E}_{R,y}$) will be non-zero.

The light from the reference arm of the AMI only contains temporal frequencies arising from field oscillations. Therefore, to good approximation, the only frequency values which will yield non-zero amplitudes will be when $|\omega_{5,6}| \in \{10^{14}\text{s}^{-1}, 10^{15}\text{s}^{-1}\}$. Because of this, it is helpful to consider the sinc function with the aforementioned relevant frequencies, and with a slight algebraic modification:

$$\text{sinc}\left(\frac{10^{-6}\text{s}(10^{14}\text{s}^{-1})\overline{\Omega}_{56}}{2}\right) = \text{sinc}\left(\frac{10^8}{2}(\overline{\omega}_5 + \overline{\omega}_6)\right),$$

where $\overline{\omega}_l \equiv \omega_l / 10^{14}\text{s}^{-1}$; and so $|\overline{\omega}_l| \in \{1, 10\}$. In this above representation it becomes clear to see, that for relevant values of $\omega$ (i.e. those which will yield non-zero amplitudes $\tilde{E}_{R,y}$), the sinc function takes the form $\text{sinc}[\beta x]$, where $\beta = 10^8/2$ and $x = \overline{\omega}_5 + \overline{\omega}_6$. Because $\text{sinc}[\beta x]$ becomes $\delta(x)/\beta = \delta(\beta x)$ as $\beta \to \infty$, the sinc function becomes a Dirac delta distribution:

$$\frac{\eta a}{2\pi c \mu_0} \int_{-\infty}^{\infty} \int_{-\infty}^{\infty} \tilde{E}_{R,y}(\omega_5) \tilde{E}_{R,y}(\omega_6) e^{i\xi\left(\frac{\omega_5}{c}+\frac{\omega_6}{c}\right)} e^{-it\Omega_{56}} \delta\left(\frac{T_d}{2}(\omega_5 + \omega_6)\right) d\omega_5 d\omega_6.$$

Then with integration over $\omega_5$, and demanding $\omega_5 = -\omega_6$ from the Dirac delta distribution, the above straightforwardly reduces to

$$\frac{\eta a}{2\pi c \mu_0} \int_{-\infty}^{\infty} \tilde{E}_{R,y}(-\omega_6) \tilde{E}_{R,y}(\omega_6) d\omega_6 = \frac{\eta a}{2\pi c \mu_0} \int_{-\infty}^{\infty} |\tilde{E}_{R,y}(\omega_6)|^2 d\omega_6,$$

which matches the $\mathcal{E}_{RR}$ term in main text equation (7).

## Solving for the Expansion Coefficients in Main Text Equations (9) and (10)

Here we determine the expansion coefficient, $U_n(\xi)$ found in Equation (9) of the main text in terms of the fields. We start by equating expressions for $\mathcal{E}_{SR}(t,\xi)$ in the main text Equations (7) and (9). We then multiply both sides by $\frac{1}{bT_c}\cos\left(\frac{n'2\pi}{T_c}t\right)$, and integrate ($n'$, $b$ are positive integers):

$$\frac{1}{bT_c}\int_{-bT_c}^{bT_c}\sum_{n\geq 0}^{\infty} U_n(\xi)\cos\left(\frac{n2\pi}{T_c}t+\theta_n(\xi)\right)\cos\left(\frac{n'2\pi}{T_c}t\right)dt =$$

$$\frac{1}{bT_c}\int_{-bT_c}^{bT_c}\frac{\eta a}{\pi c\mu_0}\int_{-\infty}^{\infty}\int_{-\infty}^{\infty}\tilde{E}_{S,y}(\omega_3)\tilde{E}_{R,y}(\omega_4)e^{i\xi\frac{\omega_4}{c}}\operatorname{sinc}\left(\frac{T_d\Omega_{34}}{2}\right)e^{-i\Omega_{34}t}\,d\omega_3\,d\omega_4\cos\left(\frac{n'2\pi}{T_c}t\right)dt$$

Carrying the multiplication and integration through gives (while recalling $\frac{n'2\pi}{T_c} = n'\omega_c$)

$$\sum_{n\geq 0}^{\infty} U_n(\xi)\left[\frac{1}{bT_c}\int_{-bT_c}^{bT_c}\cos\left(\frac{n2\pi}{T_c}t+\theta_n(\xi)\right)\cos\left(\frac{n'2\pi}{T_c}t\right)dt\right] =$$

$$\frac{\eta a}{\pi c\mu_0}\int_{-\infty}^{\infty}\int_{-\infty}^{\infty}\tilde{E}_{S,y}(\omega_3)\tilde{E}_{R,y}(\omega_4)e^{i\xi\frac{\omega_4}{c}}\operatorname{sinc}\left(\frac{T_d\Omega_{34}}{2}\right)\left[\frac{1}{bT_c}\int_{-bT_c}^{bT_c}e^{-i\Omega_{34}t}\cos(n'\omega_c t)dt\right]d\omega_3\,d\omega_4$$

and the cosine-cosine integral on the left-hand side becomes, through orthogonality relations, for all positive integers $b$ and $n'$

$$\sum_{n\geq 0}^{\infty} U_n(\xi)\left[\delta_{n,n'}\cos(\theta_n(\xi))\right] =$$

$$\frac{\eta a}{\pi c\mu_0}\int_{-\infty}^{\infty}\int_{-\infty}^{\infty}\tilde{E}_{S,y}(\omega_3)\tilde{E}_{R,y}(\omega_4)e^{i\xi\frac{\omega_4}{c}}\operatorname{sinc}\left(\frac{T_d\Omega_{34}}{2}\right)\left[\frac{1}{2bT_c}\int_{-bT_c}^{bT_c}e^{-i\Omega_{34}t}\left(e^{in'\omega_c t}+e^{-in'\omega_c t}\right)dt\right]d\omega_3\,d\omega_4$$

while we have additionally written $\cos(n'\omega_c t)$ in terms of a sum of complex exponentials. So, the Kronecker delta on the left-hand side above requires $n \to n'$, and assuming $\cos(\theta_{n'}(\xi))$ will never be zero we bring it to the right-hand side, and isolate the arbitrary expansion coefficient $U_{n'}$:

$$U_{n'}(\xi) = \frac{\eta a}{\pi c \mu_0 \cos(\theta_{n'}(\xi))} \int_{-\infty}^{\infty} \int_{-\infty}^{\infty} \tilde{E}_{S,y}(\omega_3) \tilde{E}_{R,y}(\omega_4) e^{i\xi \frac{\omega_4}{c}} \operatorname{sinc}\left(\frac{T_d \Omega_{34}}{2}\right) \times$$

$$\left[\frac{1}{2bT_c} \int_{-bT_c}^{bT_c} \left(e^{-i(\Omega_{34}-n'\omega_c)t} + e^{-i(\Omega_{34}+n'\omega_c)t}\right) dt\right] d\omega_3 \, d\omega_4$$

Performing the integrations in the closed brackets yields:

$$U_{n'}(\xi) = \frac{\eta a}{\pi c \mu_0 \cos(\theta_{n'}(\xi))} \int_{-\infty}^{\infty} \int_{-\infty}^{\infty} \tilde{E}_{S,y}(\omega_3) \tilde{E}_{R,y}(\omega_4) e^{i\xi \frac{\omega_4}{c}} \operatorname{sinc}\left(\frac{T_d \Omega_{34}}{2}\right) \times$$

$$\{\operatorname{sinc}[bT_c(n'\omega_c - \Omega_{34})] + \operatorname{sinc}[bT_c(n'\omega_c + \Omega_{34})]\} \, d\omega_3 \, d\omega_4$$

In that the above equation is true for *any* b that is both an integer and positive, without loss of generality, we take the limit of the above equation as $b$ goes to an infinitely large positive integer. In this limit both sinc functions within the curly brackets become Dirac delta distributions (because as $\beta \to \infty$, $\operatorname{sinc}[\beta x] \to \delta(x)/\beta = \delta(\beta x)$), giving

$$U_{n'}(\xi) = \frac{\eta a}{\pi c \mu_0 \cos(\theta_{n'}(\xi))} \int_{-\infty}^{\infty} \int_{-\infty}^{\infty} \tilde{E}_{S,y}(\omega_3) \tilde{E}_{R,y}(\omega_4) e^{i\xi \frac{\omega_4}{c}} \operatorname{sinc}\left(\frac{T_d \Omega_{34}}{2}\right) \times$$

$$[\delta(bT_c(n'\omega_c - \Omega_{34})) + \delta(bT_c(n'\omega_c + \Omega_{34}))] \, d\omega_3 \, d\omega_4.$$

With integration over $\omega_3$, and recalling that $\Omega_{34} = \omega_3 + \omega_4$,

$$U_{n'}(\xi) = \frac{\eta a \operatorname{sinc}\left(\frac{T_d n' \omega_c}{2}\right)}{\pi c \mu_0 \cos(\theta_{n'}(\xi))} \int_{-\infty}^{\infty} [\tilde{E}_{S,y}(-\omega_4 + n'\omega_c) + \tilde{E}_{S,y}(-\omega_4 - n'\omega_c)] \tilde{E}_{R,y}(\omega_4) e^{i\xi \frac{\omega_4}{c}} d\omega_4.$$

By using the reality condition (since $E_{s,y}(t)$ is real-valued then $\tilde{E}_{S,y}(-\omega) = \tilde{E}_{S,y}^*(\omega)$), allowing the integration variable $\omega_4$ to be rewritten as $\omega$, and letting random index $n' \to n$, the analytical solution for expansion coefficients $U_n(\xi)$ found in main text Equation (9) is

$$U_n(\xi) = \frac{\eta a \operatorname{sinc}\left(\frac{T_d n \omega_c}{2}\right)}{\pi c \mu_0 \cos(\theta_n(\xi))} \int_{-\infty}^{\infty} [\tilde{E}_{S,y}^*(\omega - n\omega_c) + \tilde{E}_{S,y}^*(\omega + n\omega_c)] \tilde{E}_{R,y}(\omega) e^{i\xi \frac{\omega}{c}} d\omega.$$

As a final, but important note, in the most recent integration step where we integrated over $\omega_3$, there was a choice to integrate over either $\omega_3$ or $\omega_4$. However, if we integrated over $\omega_4$, instead of $\omega_3$, we would have forced an a-physical result. In particular, The FT of the reference field would have functional dependence on the cantilever frequency $n\omega_c$, which cannot occur.

## Lock-in Amplification Theory Relevant to Nano-FTIR

Suppose we have a noisy signal with the following functional form:

$$N(t) = \alpha + n(t) + f(t) \tag{S1}$$

where $\alpha$ is a constant background, and

$$N, \alpha, n, f \in \mathbb{R} \; \forall \; t$$

We additionally know that $f(t)$ is periodic such that $f(t) = f(t + T_c)$. Then, we know

$$f(t) = \sum_{n=0}^{\infty} U_n \cos(n\omega_c t + \theta_n) \tag{S2}$$

where $\omega_c = \frac{2\pi}{T_c} = 2\pi f_c$, $nf_c$ for $n > 0$ is called the "$n^{th}$ harmonic frequency", and $U_n, \theta_n \in \mathbb{R}$. Note that when $n = 1$, $nf_c$ can also be referred to as the "fundamental" or "fundamental frequency". Additionally, note that often times $n\omega_c$ is *also* referred to as the "$n^{th}$ harmonic frequency," (though it would be more accurate to refer to $n\omega_c$ as the "nth *angular* harmonic frequency"). In any case, as before, all quantities (coefficients and phase constants) are real-valued. Then, without loss of generality, equation (S1) can be cast as:

$$N(t) = \alpha + n(t) + \sum_{n=0}^{\infty} U_n \cos(n\omega_c t + \theta_n) \tag{S3}$$

Now, if one is interested in determining/extracting information on $f(t)$ out of the noisy signal $N(t)$, lock-in amplification can be used. A lock-in amplifier uses robust hardware and software methods to, in effect, perform mathematical operations on a signal (like in equation (S3)) in order to determine coefficients $U_n$ and phases $\theta_n$. In this work, our purpose is not to go into a lengthy and detailed description of the electrical engineering that enables such signal manipulation, but rather to motivate, establish, and detail, the effective mathematics that such devices carry out.

Because $f(t)$ is expressed as a summation of cosine functions with different integer multiples of the fundamental frequency, $\omega_c = 2\pi f_c$, the lock-in amplifier "demodulates" the signal at chosen harmonics of interest. In general, the demodulation process consists of the lock-in amplifier conducting two operations on $N(t)$, per harmonic. The operations are simply to time average the product of the signal and a sinusoid (in our case, the cantilever drive signal). Below we give the precise operations in a general way for an arbitrary $n^{th}$ harmonic:

<u>Operation 1</u>: $\langle N(t)\cos(n\omega_c t) \rangle_{T_L} = \frac{1}{T_L} \int_{-T_L/2}^{T_L/2} N(t)\cos(n\omega_c t) dt \in \mathbb{R}$

<u>Operation 2</u>: $-\langle N(t)\sin(n\omega_c t) \rangle_{T_L} = -\frac{1}{T_L} \int_{-T_L/2}^{T_L/2} N(t)\sin(n\omega_c t) dt \in \mathbb{R}$

$$\tag{S4}$$

We want to stress that both the above operations are real-valued operations being conducted on a real function, $N(t)$. And, as such, there output is also real, as we have indicated above. However, it is extremely common to map/transform these real numbers to the complex plane by *defining/creating* a *new* complex number, say, $z_n$, who's real (imaginary) part is given by operation 1 (operation 2). Thus, one can construct

$$z_n \equiv A_n e^{i\Theta} = X_n + i\, Y_n = A_n \cos\Theta_n + i A_n \sin\Theta_n$$

Where, as mentioned before,

$$Re\{z_n\} \equiv X_n = A_n \cos\Theta_n = \frac{1}{T_L} \int_{-\frac{T_L}{2}}^{\frac{T_L}{2}} N(t)\cos(n\omega_c t)dt \in \mathbb{R}$$

and

$$Im\{z_n\} \equiv Y_n = A_n \sin\Theta_n = -\frac{1}{T_L} \int_{-\frac{T_L}{2}}^{\frac{T_L}{2}} N(t)\sin(n\omega_c t)dt \in \mathbb{R}$$

(S5)

Then the amplitude and the phase of this defined complex number are given by

$$|z_n| = A_n = \sqrt{X_n^2 + Y_n^2} = \sqrt{(A_n \cos\Theta_n)^2 + (A_n \sin\Theta_n)^2}$$

and

$$\Theta_n = \arctan\left(\frac{Y_n}{X_n}\right)$$

(S6)

As likely clear by now, it is possible to think of the two operations defined in (S4) that are actually occurring, as a *single* complex operation:

<u>Complex Operation 1</u>: $\langle N(t)e^{-in'\omega_c t}\rangle_{T_L} = \frac{1}{T_L} \int_{-T_L/2}^{T_L/2} N(t)e^{-in'\omega_c t}dt \equiv z_{n'}$

(S7)

For this reason, most literature surrounding lock-in amplifiers will make mention to a real part, imaginary part, amplitude, and phase. These values are those of this *as defined complex number* described above. For example, Zurich Instruments (maker of the Lock-in used at beamline 5.4 of the ALS) enables the user to output any of these four demodulated quantities. For a given harmonic, the following four outputs related to $z_{n'}$ can, in principle, be output: $X_{n'}, Y_{n'}, A_{n'}$, and $\Theta_{n'}$.

Now that we have covered all the notation, we will proceed with carrying out the mathematical operation/s in order to investigate the lock-in result, and in particular, emphasize how the result relates to the *original* periodic function $f(t)$ that we had interest in. Performing the effective complex operation 1 given in equation (S7) we find (for $n' \geq 1$)

$$z_{n'} = \frac{1}{T_L} \int_{-\frac{T_L}{2}}^{\frac{T_L}{2}} \left( \alpha + n(t) + \frac{1}{2}U_0\left[e^{i\theta_n} + e^{-i\theta_n}\right] + \frac{1}{2}\sum_{n=1}^{\infty} U_n\left[e^{i(n\omega_c t + \theta_n)} + e^{-i(n\omega_c t + \theta_n)}\right] \right) e^{-in'\omega_c t} dt$$

$$z_{n'} = \frac{1}{T_L} \int_{-\frac{T_L}{2}}^{\frac{T_L}{2}} \left( \alpha e^{-in'\omega_c t} + n(t)e^{-in'\omega_c t} + \frac{1}{2}U_0\left[e^{i(\theta_n - n'\omega_c t)} + e^{-i(\theta_n + n'\omega_c t)}\right] \right.$$
$$\left. + \frac{1}{2}\sum_{n=1}^{\infty} U_n\left[e^{i\theta_n}e^{i(n-n')\omega_c t} + e^{-i\theta_n}e^{-i(n+n')\omega_c t}\right] \right) dt$$

In that the first and third integrals above are just the time averages of sinusoids with constant coefficients, and then divided by a positive number, they become negligibly small. As an aside, it should also be noted before moving on, that in our case the cantilever oscillation period is $T_C \sim 10^{-5}$s, while the lock-in period is $T_L = \sim 10^{-4}$s. So, at the fundamental ($n' = 1$), the time average is occurring over about 10 cantilever oscillations ($\frac{1}{T_C}T_L$). At higher harmonics, the time average is occurring over *more oscillations*. Moving on we have

$$z_{n'} = \frac{1}{T_L}\int_{-\frac{T_L}{2}}^{\frac{T_L}{2}} n(t)e^{-in'\omega_c t}dt + \frac{1}{2}\sum_{n=1}^{\infty} U_n\left[e^{i\theta_n}\left\{\frac{1}{T_L}\int_{-\frac{T_L}{2}}^{\frac{T_L}{2}} e^{i(n-n')\omega_c t}dt\right\} + e^{-i\theta_n}\left\{\frac{1}{T_L}\int_{-\frac{T_L}{2}}^{\frac{T_L}{2}} e^{-i(n+n')\omega_c t}dt\right\}\right]$$

Analytically evaluating the two integrals on the right-hand side in the curly brackets gives

$$z_{n'} = \frac{1}{T_L}\int_{-\frac{T_L}{2}}^{\frac{T_L}{2}} n(t)e^{-in'\omega_c t}dt + \frac{1}{2}\sum_{n=1}^{\infty} U_n\left[e^{i\theta_n}\left\{\frac{\mathrm{Sin}\left(\frac{T_L(n-n')\omega_c}{2}\right)}{\frac{T_L(n-n')\omega_c}{2}}\right\} + e^{-i\theta_n}\left\{\frac{\mathrm{Sin}\left(\frac{T_L(n+n')\omega_c}{2}\right)}{\frac{T_L(n+n')\omega_c}{2}}\right\}\right]$$

For realistic values of our system, the sinc functions effectively become Kronecker delta functions

$$z_{n'} = \frac{1}{T_L}\int_{-\frac{T_L}{2}}^{\frac{T_L}{2}} n(t)e^{-in'\omega_c t}dt + \frac{1}{2}\sum_{n=1}^{\infty} U_n\left[e^{i\theta_n}\delta_{n,n'} + e^{-i\theta_n}\delta_{-n,n'}\right]$$

Because both $n, n' > 0$ the second Kronecker delta vanishes giving.

$$z_{n'} = \frac{1}{T_L} \int_{-\frac{T_L}{2}}^{\frac{T_L}{2}} n(t) e^{-in'\omega_c t} dt + \frac{1}{2} U_{n'} e^{i\theta_{n'}}$$

Lastly, in the above, if the noise is randomized, non-monotonically increasing function, the first integral is also zero giving the following (with the prime dropped).

$$z_n = \frac{U_n e^{i\theta_n}}{2} = A_n e^{i\Theta} = X_n + i Y_n$$

So, from direct comparison it is clear that the lock-in outputs are related to $f(t) = \sum_{n=0}^{\infty} U_n(\xi) \cos(n\omega_c t + \theta_n(\xi))$ in the following way:

| Lock-in Output | Relation to $f(t)$ | If via software, $\Theta_n \to 0$ (or if $\Theta_n$ is very small) |
|---|---|---|
| $X_n(\xi)$ | $\frac{U_n(\xi)}{2} \cos(\theta_n(\xi))$ | $\frac{U_n}{2}$ |
| $Y_n(\xi)$ | $\frac{U_n(\xi)}{2} \sin(\theta_n(\xi))$ | 0 |
| $A_n(\xi)$ | $\frac{U_n(\xi)}{2}$ | $\frac{U_n}{2}$ |
| $\Theta_n(\xi)$ | $\theta_n(\xi)$ | 0 |

From the above general relations, we note that the phase $\Theta_n$ (magnitude $A_n$) of the made-up complex number $z_n$, is *exactly equivalent to(proportional to) the $n^{th}$ harmonic's phase $\theta_n$ (expansion coefficient $U_n$) of the function of interest*, $f(t)$. One or a combination of these lock-in outputs are passed onto the next stage of processing.

## Algebraic Simplifications to Derive Main Text Equation (23)

Starting just after main text (22)

$$\bar{z}_{n>1}(\nu)$$

$$= \frac{\dfrac{-\omega_n^{+2}\mu_0 e^{\frac{i\omega_n^+ R}{c}}}{4\pi R}(-\tilde{p}_x^{t\&sm}(\omega_n^+)\cos[\theta_0] + \tilde{p}_z^{t\&sm}(\omega_n^+)\sin[\theta_0])g(\omega_n^+) + \dfrac{-\omega_n^{-2}\mu_0 e^{\frac{i\omega_n^- R}{c}}}{4\pi R}(-\tilde{p}_x^{t\&sm}(\omega_n^-)\cos[\theta_0] + \tilde{p}_z^{t\&sm}(\omega_n^-)\sin[\theta_0])g(\omega_n^-)}{\dfrac{-\omega_n^{+2}\mu_0 e^{\frac{i\omega_n^+ R}{c}}}{4\pi R}(-\tilde{p}_x^{t\&rm}(\omega_n^+)\cos[\theta_0] + \tilde{p}_z^{t\&rm}(\omega_n^+)\sin[\theta_0])g(\omega_n^+) + \dfrac{-\omega_n^{-2}\mu_0 e^{\frac{i\omega_n^- R}{c}}}{4\pi R}(-\tilde{p}_x^{t\&rm}(\omega_n^-)\cos[\theta_0] + \tilde{p}_z^{t\&rm}(\omega_n^-)\sin[\theta_0])g(\omega_n^-)}$$

Now conducting some algebraic cancelations and division by $\sin[\theta_0]$ we find,

$$= \frac{\omega_n^{+2} e^{\frac{i\omega_n^+ R}{c}} g(\omega_n^+)(-\tilde{p}_x^{t\&sm}(\omega_n^+)\cot[\theta_0] + \tilde{p}_z^{t\&sm}(\omega_n^+)) + \omega_n^{-2} e^{\frac{i\omega_n^- R}{c}} g(\omega_n^-)(-\tilde{p}_x^{t\&sm}(\omega_n^-)\cot[\theta_0] + \tilde{p}_z^{t\&sm}(\omega_n^-))}{\omega_n^{+2} e^{\frac{i\omega_n^+ R}{c}} g(\omega_n^+)(-\tilde{p}_x^{t\&rm}(\omega_n^+)\cot[\theta_0] + \tilde{p}_z^{t\&rm}(\omega_n^+)) + \omega_n^{-2} e^{\frac{i\omega_n^- R}{c}} g(\omega_n^-)(-\tilde{p}_x^{t\&rm}(\omega_n^-)\cot[\theta_0] + \tilde{p}_z^{t\&rm}(\omega_n^-))}$$

Now we define

$$\mathcal{D}^j(\omega) = e^{\frac{i\omega R}{c}} g(\omega)\mathcal{M}^j(\omega) = e^{\frac{i\omega R}{c}} g(\omega)\left(-\tilde{p}_x^j(\omega)\cot[\theta_0] + \tilde{p}_z^j(\omega)\right) \quad , \quad j = t\&sm \ OR \ t\&rm$$

and

$$\mathcal{M}^j(\omega) = -\tilde{p}_x^j(\omega)\cot[\theta_0] + \tilde{p}_z^j(\omega) \quad , \quad j = t\&sm \ OR \ t\&rm$$

Using this in the above gives…

$$= \frac{\omega_n^{+2}\mathcal{D}^{t\&sm}(\omega_n^+) + \omega_n^{-2}\mathcal{D}^{t\&sm}(\omega_n^-)}{\omega_n^{+2}\mathcal{D}^{t\&rm}(\omega_n^+) + \omega_n^{-2}\mathcal{D}^{t\&rm}(\omega_n^-)}$$

Continuing on,

$$= \frac{(\omega^2 + 2n\omega_c\omega + n^2\omega_c^2)\mathcal{D}^{t\&sm}(\omega_n^+) + (\omega^2 - 2n\omega_c\omega + n^2\omega_c^2)\mathcal{D}^{t\&sm}(\omega_n^-)}{(\omega^2 + 2n\omega_c\omega + n^2\omega_c^2)\mathcal{D}^{t\&rm}(\omega_n^+) + (\omega^2 - 2n\omega_c\omega + n^2\omega_c^2)\mathcal{D}^{t\&rm}(\omega_n^-)}$$

$$= \frac{\omega^2\mathcal{D}^{t\&sm}(\omega_n^+) + 2n\omega_c\omega\mathcal{D}^{t\&sm}(\omega_n^+) + n^2\omega_c^2\mathcal{D}^{t\&sm}(\omega_n^+) + \omega^2\mathcal{D}^{t\&sm}(\omega_n^-) - 2n\omega_c\omega\mathcal{D}^{t\&sm}(\omega_n^-) + n^2\omega_c^2\mathcal{D}^{t\&sm}(\omega_n^-)}{\omega^2\mathcal{D}^{t\&rm}(\omega_n^+) + 2n\omega_c\omega\mathcal{D}^{t\&rm}(\omega_n^+) + n^2\omega_c^2\mathcal{D}^{t\&rm}(\omega_n^+) + \omega^2\mathcal{D}^{t\&rm}(\omega_n^-) - 2n\omega_c\omega\mathcal{D}^{t\&rm}(\omega_n^-) + n^2\omega_c^2\mathcal{D}^{t\&rm}(\omega_n^-)}$$

$$= \frac{\omega^2(\mathcal{D}^{t\&sm}(\omega_n^+) + \mathcal{D}^{t\&sm}(\omega_n^-)) + 2n\omega_c\omega(\mathcal{D}^{t\&sm}(\omega_n^+) - \mathcal{D}^{t\&sm}(\omega_n^-)) + n^2\omega_c^2(\mathcal{D}^{t\&sm}(\omega_n^+) + \mathcal{D}^{t\&sm}(\omega_n^-))}{\omega^2(\mathcal{D}^{t\&rm}(\omega_n^+) + \mathcal{D}^{t\&rm}(\omega_n^-)) + 2n\omega_c\omega(\mathcal{D}^{t\&rm}(\omega_n^+) - \mathcal{D}^{t\&rm}(\omega_n^-)) + n^2\omega_c^2(\mathcal{D}^{t\&rm}(\omega_n^+) + \mathcal{D}^{t\&rm}(\omega_n^-))}$$

Further, if we multiply top and bottom of the previous by $1/(2n\omega_c\omega^2)$ we have….

$$= \frac{\left\{\frac{(\mathcal{D}^{t\&sm}(\omega_n^+) + \mathcal{D}^{t\&sm}(\omega_n^-))}{2n\omega_c}\right\} + 2\left(\frac{n\omega_c}{\omega}\right)\left\{\frac{(\mathcal{D}^{t\&sm}(\omega_n^+) - \mathcal{D}^{t\&sm}(\omega_n^-))}{2n\omega_c}\right\} + \left(\frac{n\omega_c}{\omega}\right)^2\left\{\frac{(\mathcal{D}^{t\&sm}(\omega_n^+) + \mathcal{D}^{t\&sm}(\omega_n^-))}{2n\omega_c}\right\}}{\left\{\frac{(\mathcal{D}^{t\&rm}(\omega_n^+) + \mathcal{D}^{t\&rm}(\omega_n^-))}{2n\omega_c}\right\} + 2\left(\frac{n\omega_c}{\omega}\right)\left\{\frac{(\mathcal{D}^{t\&rm}(\omega_n^+) - \mathcal{D}^{t\&rm}(\omega_n^-))}{2n\omega_c}\right\} + \left(\frac{n\omega_c}{\omega}\right)^2\left\{\frac{(\mathcal{D}^{t\&rm}(\omega_n^+) + \mathcal{D}^{t\&rm}(\omega_n^-))}{2n\omega_c}\right\}}$$

Then, if we define

$$\Delta_\pm^j = \frac{\mathcal{D}^j(\omega_n^+) \pm \mathcal{D}^j(\omega_n^-)}{2n\omega_c}$$

the boxed expression for $\bar{z}_{n>1}(\nu)$ reduces to main text Equation (23):

$$\bar{z}_{n>1}(\nu) = \frac{\left(\frac{n\omega_c}{\omega}\right)^0 \Delta_+^{t\&sm} + \left(\frac{n\omega_c}{\omega}\right)^1 2\Delta_-^{t\&sm} + \left(\frac{n\omega_c}{\omega}\right)^2 \Delta_+^{t\&sm}}{\left(\frac{n\omega_c}{\omega}\right)^0 \Delta_+^{t\&rm} + \left(\frac{n\omega_c}{\omega}\right)^1 2\Delta_-^{t\&rm} + \left(\frac{n\omega_c}{\omega}\right)^2 \Delta_+^{t\&rm}}$$

As currently expressed, each of the three terms in the numerator (and denominator) are scaled by a small dimensionless multiplicative perfector, raised to various powers of $m$: $(n\omega_c/\omega)^m$. For the system at hand, $(n\omega_c/\omega) \sim \mathcal{O}(10^{-8}) \ll 1$ for all relevant optical (angular) frequencies. Therefore, the power $m$ of the perfector defines each terms' degree of smallness: zeroth order in smallness ($m = 0$), first order in smallness ($m = 1$), and second order in smallness ($m = 2$).

## General Dipole Moment Relations

The most general equation for an electric dipole is:[1]

$$\boldsymbol{p}(t) \equiv \int_V \boldsymbol{x}'(t) \rho_{tot}(\boldsymbol{x}',t) d^3 x' \tag{S8}$$

where the $\rho_{tot}$ is the *total* charge (i.e. bound and free), and the spatial volume integral is over the *arbitrary* charge density. Then by using $\epsilon_0 \boldsymbol{\nabla} \cdot \boldsymbol{E} = \rho_{tot}$, followed by $\boldsymbol{E} = \epsilon_0^{-1}(\boldsymbol{D} - \boldsymbol{P})$, (S8) becomes,

$$\boldsymbol{p}(t) = \int_V \boldsymbol{x}'(t) \boldsymbol{\nabla} \cdot \boldsymbol{D}\, d^3 x' - \int_V \boldsymbol{x}'(t) \boldsymbol{\nabla} \cdot \boldsymbol{P} d^3 x'. \tag{S9}$$

With the help of $\boldsymbol{\nabla} \cdot \boldsymbol{D} = \rho_{free}$ and Jackson (9.14),[1] the above becomes,

$$\boldsymbol{p}(t) = \int_V \boldsymbol{x}'(t)\, \rho_{free}(\boldsymbol{x}',t)\, d^3 x' + \int_V \boldsymbol{P}\, d^3 x'. \tag{S10}$$

Then, using $\boldsymbol{P} = \epsilon_0 \hat{\chi}\, \boldsymbol{E}$ in (S10), we arrive at:

$$\boldsymbol{p}(t) = \int_V \boldsymbol{x}'(t)\, \rho_{free}(\boldsymbol{x}',t)\, d^3 x' + \int_V \epsilon_0 \hat{\chi}\, \boldsymbol{E}\, d^3 x'. \tag{S11}$$

We stress that "hat" notations overtop Greek characters in this section are to notate linear integral operators. The left term gives the free charge contributing to the dipole moment and the right term gives the bound charge contributing to the dipole moment. It follows that the second time derivative of a dipole moment from a generalized charge distribution is

$$\ddot{\boldsymbol{p}}(t) = \int_V \partial_t^2 \boldsymbol{x}'(t)\, \rho_{free}(\boldsymbol{x}',t)\, d^3 x' + \epsilon_0 \int_V \partial_t^2 \hat{\chi}\, \boldsymbol{E}\, d^3 x'. \tag{S12}$$

Using the continuity equation, $-\boldsymbol{\nabla} \cdot \boldsymbol{j}_{free} = \partial_t \rho_{free}$, in (S12),

$$\ddot{\boldsymbol{p}}(t) = -\int_V \partial_t \boldsymbol{x}'(t)\, \boldsymbol{\nabla} \cdot \boldsymbol{j}_{free}\, d^3 x' + \epsilon_0 \int_V \partial_t^2 \hat{\chi}\, \boldsymbol{E}\, d^3 x'. \tag{S13}$$

Then we use the constitutive relation, $\boldsymbol{j}_{free} = \hat{\sigma}\, \boldsymbol{E}$ in (S13) and find

$$\ddot{\boldsymbol{p}}(t) = -\partial_t \hat{\sigma} \int_V \boldsymbol{x}'(t)\, \boldsymbol{\nabla} \cdot \boldsymbol{E}\, d^3 x' + \epsilon_0 \int_V \partial_t^2 \hat{\chi}\, \boldsymbol{E}\, d^3 x'. \tag{S14}$$

Again, using Jackson (9.14)[1] on the first term on the right-hand side of (S14)

$$\ddot{\boldsymbol{p}}(t) = \partial_t \hat{\sigma} \int_V \boldsymbol{E}\, d^3 x' + \epsilon_0 \int_V \partial_t^2 \hat{\chi}\, \boldsymbol{E}\, d^3 x'. \tag{S15}$$

Thus,

$$\ddot{\boldsymbol{p}}(t) = (\partial_t \hat{\sigma} + \epsilon_0 \partial_t^2 \hat{\chi}) \int_V \boldsymbol{E}\, d^3x' \tag{S16}$$

and,

$$\ddot{p}_3(t)\, \boldsymbol{e_3} = (\partial_t \hat{\sigma} + \epsilon_0 \partial_t^2 \hat{\chi}) \int_V E_3 d^3x'\, \boldsymbol{e_3}. \tag{S17}$$

Now, the FT of the second time derivative of a generalized dipole moment is then,

$$\mathfrak{F}_{t \to \omega} \ddot{\boldsymbol{p}}(t) = (-i\omega\, \tilde{\sigma} - \epsilon_0 \omega^2 \tilde{\chi}) \int_V \tilde{\boldsymbol{E}}\, d^3x' \tag{S18}$$

But it is also known that $\tilde{\epsilon}_r(\omega) = 1 + \tilde{\chi} + i\frac{\tilde{\sigma}}{\varepsilon_0 \omega}$ and therefore $\tilde{\epsilon}_r(\omega) - 1 = i\frac{\tilde{\sigma}}{\varepsilon_0 \omega} + \tilde{\chi}$, and so,

$$-\epsilon_0 \omega^2 (\tilde{\epsilon}_r(\omega) - 1) = -i\omega \tilde{\sigma} - \epsilon_0 \omega^2 \tilde{\chi} \tag{S19}$$

So, (S19) into (S18)

$$\mathfrak{F}_{t \to \omega} \ddot{\boldsymbol{p}}(t) = -\epsilon_0 \omega^2 (\tilde{\epsilon}_r(\omega) - 1) \int_V \tilde{\boldsymbol{E}}\, d^3x' \tag{S20}$$

We also know that $\mathfrak{F}_{t \to \omega} \ddot{\boldsymbol{p}}(t) = -\omega^2 \tilde{\boldsymbol{p}}(\omega)$ therefore, equating this and (S20) we finally arrive at the expression given just above equation (26) in the main text

$$\boxed{\tilde{\boldsymbol{p}}(\omega) = \epsilon_0 (\tilde{\epsilon}_r(\omega) - 1) \int_V \tilde{\boldsymbol{E}}\, d^3x'} \tag{S21}$$

## Basic Model for Generalized Dipole Radiation from the Tip and Sample

Starting from main text Equation (26),

$$\bar{z}_{n>1}(\nu) \approx \frac{\tilde{p}_z^{t/sm}(\omega) + \tilde{p}_z^{sm}(\omega)}{\tilde{p}_z^{t/rm}(\omega) + \tilde{p}_z^{rm}(\omega)}, \tag{S22}$$

and from equation (S21),

$$\tilde{p}_z = \epsilon_0(\tilde{\epsilon}_r - 1)\xi_V, \quad \text{where} \quad \xi_V = \int_V \tilde{E}_z(\omega)\, dV. \tag{S23}$$

Putting these into (S22) gives

$$\bar{z}_{n>1}(\nu) \approx \frac{(\tilde{\epsilon}_r^t - 1)\xi_V^{t/sm} + (\tilde{\epsilon}_r^{sm} - 1)\xi_V^{sm}}{(\tilde{\epsilon}_r^t - 1)\xi_V^{t/rm} + (\tilde{\epsilon}_r^{rm} - 1)\xi_V^{rm}}, \tag{S24}$$

where, as in the main text, the superscript notation $t/sm$ serves to indicate "tip adjacent to sample material" and similarly, $t/rm$ indicates "tip adjacent to reference material." Because the probe tip is purposefully a strong metallic conductor we choose to divide the top and bottom of (S24) by $(\tilde{\epsilon}_r^t - 1)$:

$$\bar{z}_{n>1}(\nu) \approx \frac{\xi_V^{t/sm} + \left(\frac{\tilde{\epsilon}_r^{sm} - 1}{\tilde{\epsilon}_r^t - 1}\right)\xi_V^{sm}}{\xi_V^{t/rm} + \left(\frac{\tilde{\epsilon}_r^{rm} - 1}{\tilde{\epsilon}_r^t - 1}\right)\xi_V^{rm}}. \tag{S25}$$

But now, both $\left(\frac{\tilde{\epsilon}_r^{sm}-1}{\tilde{\epsilon}_r^t-1}\right)$ and $\left(\frac{\tilde{\epsilon}_r^{rm}-1}{\tilde{\epsilon}_r^t-1}\right)$ will be much less than one! Unless the sample or reference material is equally conductive as the probe. So, for cases of a typical reference materials such as Si, and considering a non-metallic sample material, then (S25) can be approximated as:

$$\bar{z}_{n>1}(\nu) \approx \frac{\xi_{tV}^{t/sm}}{\xi_{tV}^{t/rm}} = \frac{\int_{tV} \tilde{E}_{z,in}^{t/sm}(\omega)\, dV}{\int_{tV} \tilde{E}_{z,in}^{t/rm}(\omega)\, dV}. \tag{S26}$$

At this stage, we approximate the tip as a polarizable sphere in an external field where the wavelength of the light is much larger than the characteristic size of the sphere ($\lambda \gg \ell$). Then we know that the field inside the sphere is constant in space, and the following relation holds:[1]

$$\tilde{E}_{z,in}^j(\omega) = \frac{3}{\tilde{\epsilon}_r^t + 2} \tilde{E}_{z,out}^j = \frac{3}{\tilde{\epsilon}_r^t + 2}\left(\tilde{E}_{z,out}^{Inc} + \tilde{E}_{z,out}^{NF,(sm\ or\ rm)}\right). \tag{S27}$$

Now $\tilde{E}_{z,out}^{Inc}$ is included here to mention initial light source, but we stress that contributions to the radiation field from this term would have already vanished because of the lock-in removing it due to lack of periodicity in $T_c$. Therefore, we drop that term and insert the remaining parts of (S27) into (S26):

$$\bar{z}_{n>1}(\nu) \approx \frac{\frac{3}{\tilde{\epsilon}_r^t + 2} \tilde{E}_{z,out}^{NF,sm} \int_{tV} dV}{\frac{3}{\tilde{\epsilon}_r^t + 2} \tilde{E}_{z,out}^{NF,rm} \int_{tV} dV} = \frac{\tilde{E}_{z,out}^{NF,sm}}{\tilde{E}_{z,out}^{NF,rm}}. \quad (S28)$$

Now all that remains is to write a reasonable expression for the electric field produced from the sample (and reference) material, very close to the sample (or reference) materials' surface. Once this is had, insert it into (S28), and simplify appropriately.

For this task, we recall that a dipole moment, $\tilde{p}_z^{sm}$ ($\tilde{p}_z^{rm}$) is induced in the sample (reference) material. This dipole is induced by both the original incident light, and the electric near-field emanating from the probe tip. As a reminder, the near-field periodically penetrates a nanoscopic volume of the material below it, as the probe's cantilever dithers, and is the one we are concerned with writing an expression for (as contributions from the incident field will not directly contribute because of the lock-in).

The dipole will produce an electric field far away from it, via radiation, as discussed at length in the main text already. That said, the dipole will also produce an electric field very close to it – an electric near-field. This is the field we are concerned with modeling. The z-component of near-field produced by an arbitrary $\tilde{p}_z$ along the dipole axis and outside the charge distribution will be[1]

$$\tilde{E}_z^{NF,\tilde{p}_z} = \frac{-2\tilde{p}_z}{4\pi\epsilon_0 D^3}. \quad (S29)$$

Using (S29) in (S28) yields:

$$\bar{z}_{n>1}(\nu) \approx \frac{\frac{-2\tilde{p}_z^{sm}}{4\pi\epsilon_0 D^3}}{\frac{-2\tilde{p}_z^{rm}}{4\pi\epsilon_0 D^3}} = \frac{\tilde{p}_z^{sm}}{\tilde{p}_z^{rm}}. \quad (S30)$$

Nearing the end of the derivation, we use the familiar relationship (S23) one last time, and (S30) becomes:

$$\bar{z}_{n>1}(\nu) \approx \frac{(\tilde{\epsilon}_r^{sm} - 1)\xi_V^{sm}}{(\tilde{\epsilon}_r^{rm} - 1)\xi_V^{rm}}. \quad (S31)$$

In a way, (S31) if fairly remarkable. Even though the radiation that is detected in the far-field is dominated by radiation physically originating from the probe tip, as a result of the detection scheme (primarily the lock-in), the normalized spectrum $\bar{z}_{n>1}(\nu)$ is approximately equivalent to a ratio of the FT of the dipoles physically located in the sample and reference materials. Clearly the direct evidence of near-field coupling between probe and sample is evident.

Approximating (S31) further by allowing $\xi_V^{sm}/\xi_V^{rm} \approx 1$ (an approximation that will break down at strong resonances), yields

$$\bar{z}_{n>1}(\nu) \approx \frac{\tilde{\epsilon}_r^{sm} - 1}{\tilde{\epsilon}_r^{rm} - 1}, \quad (S32)$$

which matches main text equation (27).

## Peak Shift Analysis of Kapton Data

Below we show plots of peak center shifts for three near-field quantities relative to ATR-FTIR peak centers: (i) the imaginary part of the normalized complex-valued nano-FTIR spectrum (red), (ii) the model extinction coefficient from main text equation (29) (blue), and (iii) a model extinction coefficient derived from another commonly used simple model that is the ratio of near-field reflection coefficients (green):[2, 3] $\bar{z}_{n>1} \approx \beta^{sm}/\beta^{rm}$, where $\beta^j = (\tilde{\epsilon}_r^j - 1)/(\tilde{\epsilon}_r^j + 1)$. Peak centers (the horizontal axes in Fig. S1.) were identified from the data plotted in main text Fig.8b via Lorenz model fitting. We find that for the energy range and class of materials investigated, the model extinction coefficient via main text equation (29) has the best agreement with ATR-FTIR data.

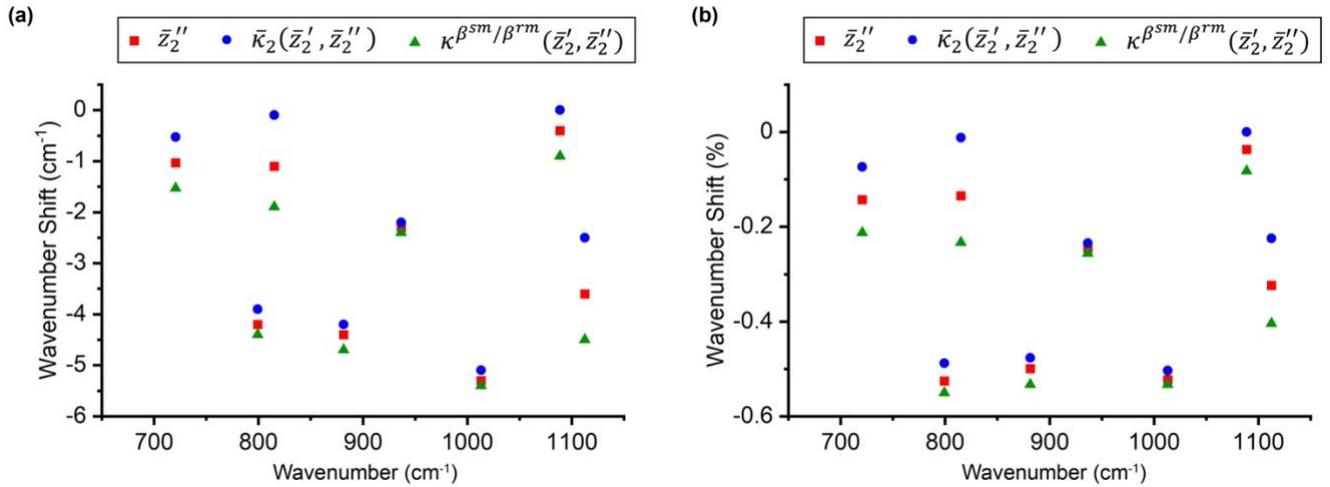

**Fig. S1.** Scatter plots of (a) absolute and (b) percent peak shifts of $\bar{z}_2''$, $\kappa_2(\bar{z}_2', \bar{z}_2'')$, and $\kappa^{\beta^{sm}/\beta^{rm}}(\bar{z}_2', \bar{z}_2'')$, relative to ATR-FTIR peak centers. The horizontal axes denotes the center of the ATR-FTIR peak. Spectra are displayed in main text Fig. 8b.

**References:**

(1) Jackson, J. D., *Classical Electrodynamics*. 3rd ed.; Wiley: 1998.
(2) Huth, F.; Govyadinov, A.; Amarie, S.; Nuansing, W.; Keilmann, F.; Hillenbrand, R. *Nano Lett.* **2012,** 12, (8), 3973-8.
(3) Govyadinov, A. A.; Amenabar, I.; Huth, F.; Carney, P. S.; Hillenbrand, R. *J. Phys. Chem. Lett.* **2013,** 4, (9), 1526-31.